\documentclass[aps,prx,longbibliography,superscriptaddress,twocolumn,showpacs]{revtex4-1}
\usepackage[utf8]{inputenc}
\usepackage[american,british]{babel}
\usepackage[T1]{fontenc}
\usepackage[pdftex]{graphicx}
\usepackage{xcolor}
\usepackage{dcolumn}
\usepackage{bm}
\usepackage{amsmath,amsthm,amssymb,mathrsfs,mathtools,amsfonts,braket}
\usepackage{verbatim}
\usepackage{epsfig}
\usepackage{color}
\usepackage{xfrac}
\usepackage{geometry}
\usepackage{hyperref}
\usepackage{bbold}
\usepackage{tensor}
\usepackage{ulem}

\DeclareMathOperator{\Tr}{Tr}
\DeclareMathOperator{\sign}{sgn}

\renewcommand{\Re}{\mathfrak{Re}}

\renewcommand{\sim}{\thicksim}
\newcommand{\numbersets}{\mathbb}
\newcommand{\Z}{\numbersets{Z}}

\newcommand{\bxi}{\boldsymbol{\xi}}

\usepackage{hyperref}
\hypersetup{
    bookmarks=false,         
    unicode=false,          
    pdftoolbar=false,        
    pdfmenubar=true,        
    pdffitwindow=false,     
    pdfstartview={FitH},    
    pdftitle={},    
    pdfauthor={Authors},     
    pdfsubject={},   
    pdfcreator={},   
    pdfproducer={}, 
    pdfnewwindow=true,      
    colorlinks=true,       
    linkcolor=black,          
    citecolor=blue,        
    filecolor=magenta,      
    urlcolor=blue           
}

\renewcommand{\Re}{\mathfrak{Re}}

\begin{document}

\title{Scaling of temporal entanglement in proximity to integrability}

\author{Alessio Lerose}
\affiliation{Department of Theoretical Physics, University of Geneva, Quai Ernest-Ansermet 30, 1205 Geneva, Switzerland}
\author{Michael Sonner}
\affiliation{Department of Theoretical Physics, University of Geneva, Quai Ernest-Ansermet 30, 1205 Geneva, Switzerland}
\author{Dmitry A. Abanin}
\affiliation{Department of Theoretical Physics, University of Geneva, Quai Ernest-Ansermet 30, 1205 Geneva, Switzerland}
\date{\today}

\begin{abstract}

Describing dynamics of quantum many-body systems is a formidable challenge due
to rapid generation of quantum entanglement between remote degrees of freedom. A
promising approach to tackle this challenge, which has been proposed recently,
is to characterize the quantum dynamics of a many-body system and its properties
as a bath via the Feynman-Vernon influence matrix (IM), which is an operator in the
space of time trajectories of local degrees of freedom. 
Physical understanding of the general scaling of the IM’s \textit{temporal entanglement} and its relation to basic dynamical properties is highly incomplete to present day. In this Article,
we analytically compute
the exact IM for a family of integrable Floquet models -- the transverse-field kicked
Ising chain -- finding a Bardeen-Cooper-Schrieffer-like ``wavefunction'' on the Schwinger-Keldysh contour with algebraically decaying correlations.
We demonstrate that the IM exhibits \textit{area-law} temporal entanglement scaling for all
parameter values. Furthermore, the entanglement pattern of the IM reveals the
system's phase diagram, exhibiting jumps across transitions between distinct
Floquet phases. Near criticality, a {non-trivial} scaling behavior
of temporal entanglement is found. The area-law temporal
entanglement allows us to efficiently describe the effects of sizeable integrability-breaking
perturbations for long evolution times by using matrix product state methods.
This work shows that tensor network methods are efficient in describing the effect of non-interacting baths
on open quantum systems, and provides a new approach to studying quantum many-body systems with
weakly broken integrability.
\end{abstract}

\maketitle

\section{Introduction} 
Understanding and classifying the non-equilibrium behavior of quantum matter represents a major endeavor in contemporary \mbox{physics~\cite{Polkovnikov-rev,AbaninRMP,NEQIntegrabilityJStat,KhemaniTimeCrystalReview,Nathan17,ScarsReview}}.
Theoretical description of quantum dynamics of  many-particle systems is a formidable challenge, as the complexity of the problem generally scales exponentially with the 
size of the system. 
This {\it exponential wall} severely limits the reach of exact numerical computations, spurring the search for analytical solutions~\cite{Calabrese06,RigolGGE,CalabreseEsslerFagottiPRL,BernardDoyonReview,BertiniGHD,DoyonGHD,NahumPRXOperatorSpreading,ChalkerPRX,Guhr1,Bertini2019}, rigorous bounds~\cite{PRBPrethermal2017,MoriPRL16_RigorousBoundHeating,ElsePrethermalTimeCrystalPRX,DeRoeckVerreet19,ElseQuasiperiodicPhasesPRX},
and approximate descriptions~\cite{Banuls09,Schollwock_ReviewMPSTimeEvolution,Verstraete_ReviewMPSTangentSpace,CarleoSciRep12,CarleoTroyerScience17,ShiAoP18}.
Furthermore, 
experimental quantum simulation platforms may give access to
certain regimes of quantum dynamics that are beyond the reach of classical
methods~\cite{QuantumSimulatorNSF,GrossBloch17}.
%
%


The ability of conventional algorithms based on matrix product states (MPS)~\cite{tebd} to simulate out-of-equilibrium quantum many-body dynamics is mainly
limited by the rapid generation of quantum entanglement between spatially
separated subsystems.
A promising idea to overcome this limitation is to develop efficient tensor-network descriptions that rely on low \textit{spatio-temporal} entanglement~\cite{Banuls09,HastingsLightCone}, arising in the space-time descriptions of quantum many-body dynamics in {multi-time} Hilbert spaces~\cite{CotlerSuperdensity,lerose2020}.
Ref.~\cite{lerose2020}, in particular, developed a self-consistent formulation of the Feynman-Vernon influence functional theory~\cite{FeynmanVernon} for periodically driven spin chains with local interactions.
The central object of this theory, the influence matrix (IM), fully encodes
the quantum noise exerted by the system on its local subsystems. The IM is a functional of the time trajectories of local degrees of freedom --- i.e., it can be viewed as a ``wavefunction'' in time rather than in space.
%
The efficiency of numerical simulations of quantum dynamics within this approach
is tied to the scaling of the maximum von Neumann entropy of the IM,
which we call here \textit{temporal entanglement} (TE) entropy, as a function of the
evolution time. 

While the growth of spatial entanglement in quantum quenches has been
extensively studied in various
regimes~\cite{CalabreseCardyReview,CalabreseFagotti08,AlbaPNAS,Kim13,Moore12,Znidaric08,we,ChalkerPRX,BertiniPRXEntanglementDU,GopalakrishnanLamacraftPRB19,NahumPRXEntanglement,KeyserlingkPRXOTOC},
much less is known about the behavior of temporal entanglement, although
pioneering investigations~\cite{Banuls09,muller2012tensor,HastingsPRAFolding}
have studied several concrete examples. Recently, it was realized that there are
several families of models where TE is small, or even vanishing, opening the
door to an efficient description of dynamical properties not accessible to other
methods. In particular, TE has been shown to vanish in certain solvable chaotic
quantum circuits characterized by dual-unitary gates~\cite{Bertini2019,Piroli2020}, due to
the fact that such systems act as perfectly Markovian baths on
themselves~\cite{lerose2020}, which corresponds to a product-state form of the IM
wavefunction.
A slow scaling of TE has been found in a spin chain exhibiting weak or suppressed thermalization~\cite{HastingsPRAFolding},
as well as in many-body localized systems with strong disorder  and weak interactions~\cite{Sonner20CharacterizingMBL}.
Furthermore, Ref.~\cite{Koblas20} effectively constructed an exact solution for an IM in the form of a finite MPS for a certain integrable quantum cellular automaton.
When such models are weakly perturbed, TE
is expected to stay relatively low, allowing one to efficiently describe the
local relaxation dynamics over long time scales and in the thermodynamic limit.
However, despite the recent progress and versatility of this approach, the basic
understanding of the behavior of temporal entanglement and its scaling with
evolution time remains highly incomplete.

The goal of the present work is to fill this gap. We unveil the scaling of TE in a class of integrable systems across quantum phase transitions,
as well as its behavior upon breaking integrability.
We consider a family of kicked interacting spin chains which includes an
integrable submanifold, where dynamics are solvable in terms of underlying
non-interacting fermionic quasiparticles.
By analytically deriving an exact expression of the IM  of these integrable
systems, we demonstrate that TE entropy displays an area-law scaling with
evolution time $t$, saturating to a finite value as $t\to\infty$.

Approaching critical lines in the phase diagram, convergence to the
asymptotic value becomes infinitely slow, leading to singular behavior of
saturated TE in the form of discontinuous jumps and associated critical scaling
behavior.
We connect this phenomenon to the singular changes occurring in the quasiparticle spectrum and with the appearance of long-lived edge coherence in the form of strong zero modes~\cite{FendleyLongCoherenceEdge}. As a byproduct, our analysis showcases the non-perturbative nature of local temporal correlations arising in circuits detuned from dual-unitary points~\cite{lerose2020,GuhrPerturbation,BertiniPerturbation,ChalkerLyapunov}.
Our results thus establish that TE serves as a sensitive probe of quantum phase transitions, even in stationary (infinite-temperature) ensembles.

As integrability gets broken by global perturbations, exact solutions are no
longer available.
The proximity to integrability suggests that the amount of TE
entropy remains parametrically low, which paves the way to efficient MPS
descriptions of the IM.
We demonstrate that the MPS approach allows to reliably compute local relaxation processes generated by global non-integrable dynamics over several tens to few hundreds of driving cycles.
Our numerical results demonstrate that integrability breaking perturbations
have qualitatively different effects on TE scaling in different regions of the phase diagram, from an apparent long-time saturation almost insensitive to the perturbation in the symmetric phase, to a parametrically slow growth in the symmetry-broken phase.
These findings suggest subtle connections between TE scaling, non-Markovianity, and edge physics in topological Floquet phases.

The rest of the paper is organized as follows: In Section~\ref{sec2}, we introduce the model and the influence matrix, deriving a convenient representation of the latter in terms of a trace over the environment degrees of freedom. In Sec.~\ref{sec3}, the IM of a kicked transverse-field Ising model is computed, and it is found that it takes the form of a Bardeen-Cooper-Schrieffer-like ``wavefunction'' in the Schwinger-Keldysh temporal domain. Using this representation, in Sec.~\ref{sec4} we numerically and analytically demonstrate the area-law scaling of TE in the limit $t\to\infty$. In Sec.~\ref{sec5}, we investigate the scaling behavior of TE near and across critical points. After that, in Sec.~\ref{sec6} we analyze the effects of integrability breaking, and develop an MPS representation of the IM away from non-interacting lines. Finally, we summarize the results of the paper and discuss directions for future research that they open in Sec.~\ref{sec7}.

\begin{figure*}
    \centering
    \includegraphics[height=0.22\textwidth]{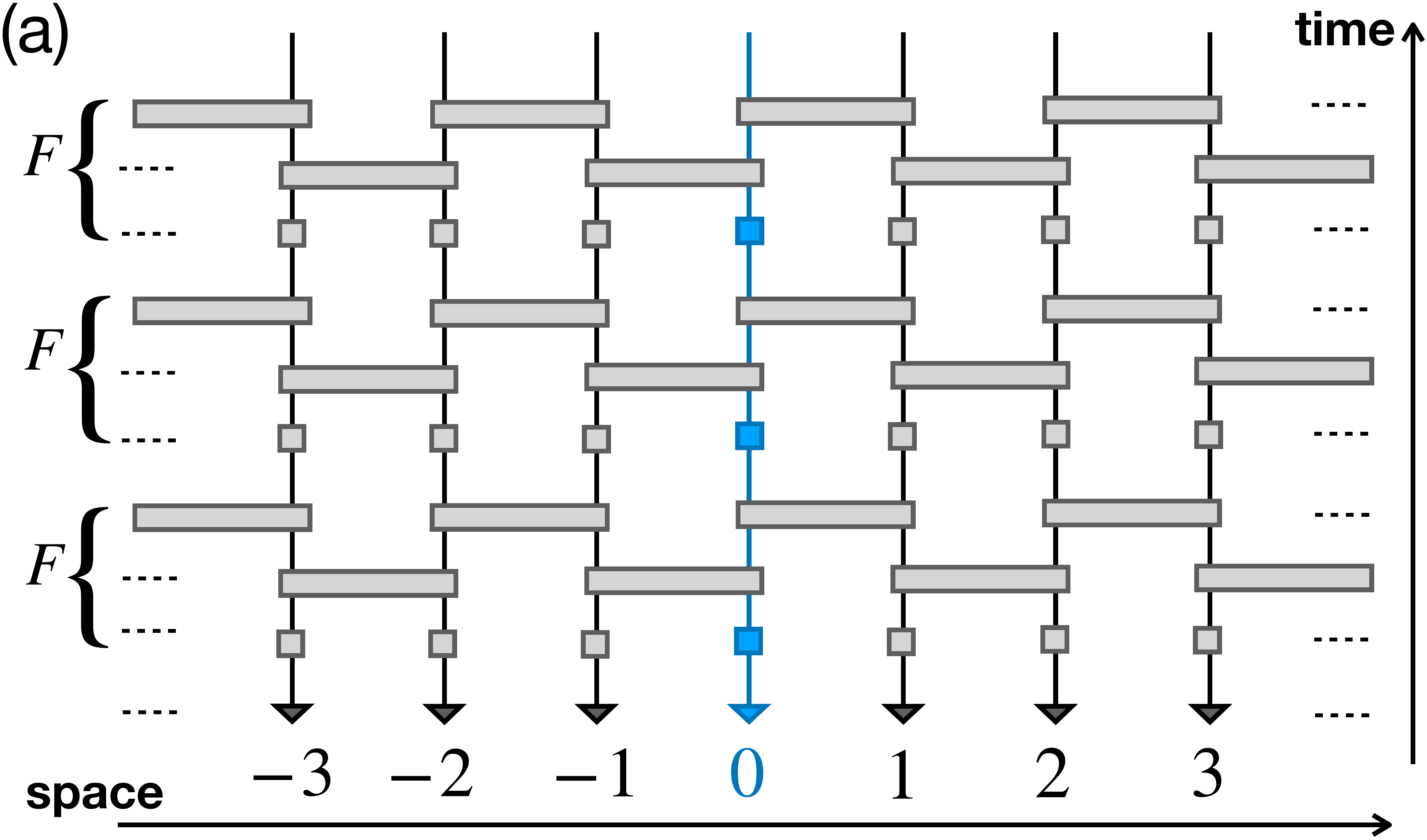}
    \hspace{0.25cm}
    \includegraphics[height=0.22\textwidth]{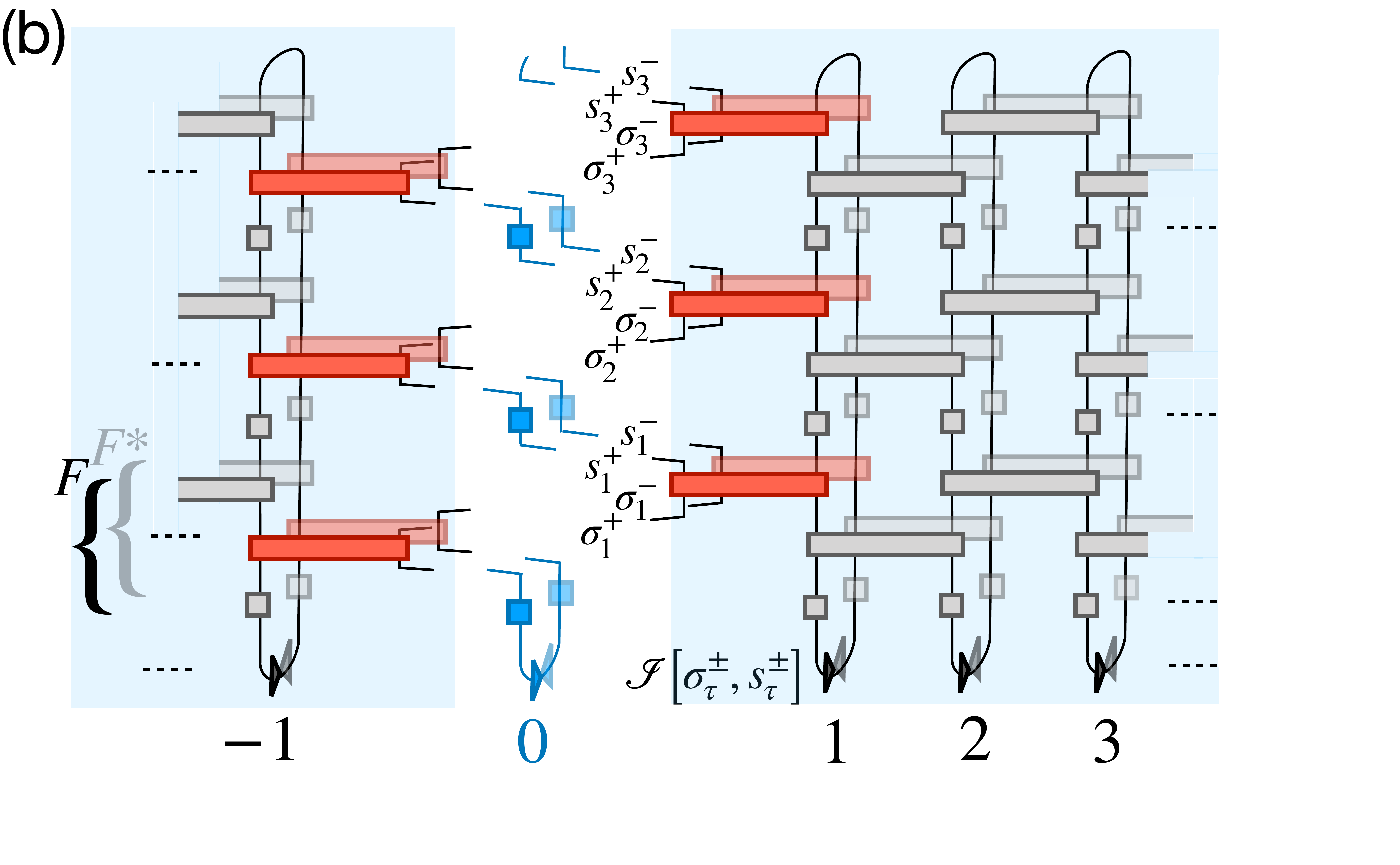}
    \hspace{0.25cm}
    \includegraphics[height=0.22\textwidth]{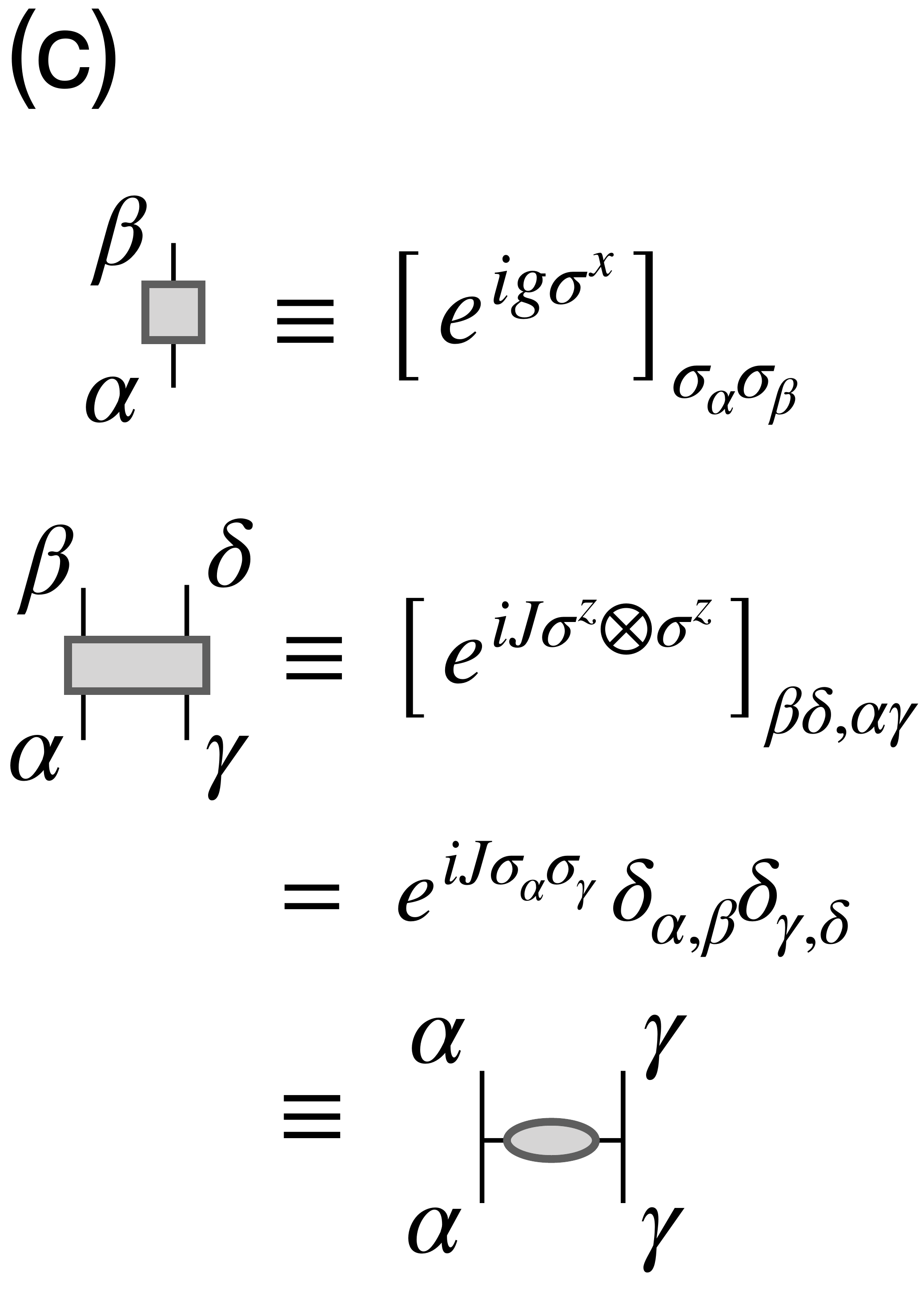}
    \caption{
    Panel (a):
    Graphical (tensor-network) representation of the circuit dynamics in Eq.~\eqref{eq_circuit}. A completely uncorrelated initial wavefunction (bottom triangles) evolves under the periodic action of single-qubit and two-qubit unitary gates forming the Floquet operator~$F$. The subsystem composed of the qubit at $j=0$, highlighted in blue, is singled out, and the qubits to its left and right are treated as its environments.
    Panel (b):
    The folded circuit represents the evolution of the initial density matrix. It can be conveniently split into the subsystem and its left and right environments. The influence matrices in Eq.~\eqref{eq_IMdef} correspond to multi-time tensors highlighted by blue shading, obtained as the contraction of the environment network (or computing the path integral over the environment trajectories).
    Arbitrary time-dependent observables and correlators pertaining to the subsystem only can be extracted by summing over the subsystem trajectories weighted by the IMs.
    The subsystem-environment decomposition of the Floquet operator $F = F_{\text{int}} F_S F_E $ is illustrated by means of the coloring (resp. red, blue, gray).
    Panel (c): Tensor notations for the gates of the transverse-field kicked Ising chain in Eq.~\eqref{eq_KIC}.
    }
    \label{fig_IM}
\end{figure*}

\section{Influence matrix for Floquet quantum circuits }\label{sec2} 
We consider Floquet unitary circuits  acting on a chain of qubits (or spins-$1/2$),
with sites indexed by $j\in\mathbb{Z}$ and computational basis
$\{\ket{\sigma}\}_{\sigma=\pm 1}$. Their dynamics are  generated by repeated
applications of the Floquet operator
\begin{equation}
\label{eq_circuit}
F =  \bigg(\prod_{b\in\mathbb{Z}} V_{2b,2b+1} \bigg)
\bigg(\prod_{b\in\mathbb{Z}} V_{2b-1,2b} \bigg)
\bigg(
\prod_{j\in\mathbb{Z}} K_{j}
\bigg)
\, .
\end{equation}
Time evolution alternates single-qubit rotations $K_j$ with even and odd local two-qubit gates $V_{j,j+1}$ periodically in time, 
as
illustrated in Fig. \ref{fig_IM}a.

We are interested in describing the dynamics of a part of the system, say, the qubit at site $j = 0$ (\textit{subsystem}),  treating the rest of the circuit to the left and to the right as \textit{environments}.
We consider a completely uncorrelated initial density matrix (DM) $\rho_0=\bigotimes_{j\in\Z} \rho_{0}^{(j)}$,
which evolves for $t$ steps, such that the DM becomes
$$
\rho_t = F^{t} \bigg(\bigotimes_{j\in\Z} \rho_{0}^{(j)}\bigg) F^{-t}.
$$
The effect of the left or right environments on the subsystem's evolution from time $\tau=0$ to $\tau=t$ is encoded in a functional of the subsystem trajectory, obtained by tracing out the environment degrees of freedom, as a function of input-output states of the subsystem at all intermediate time steps. This functional is thus a multi-time tensor, pictured in Fig.~\ref{fig_IM}b, representing a discrete-time version of the Feynman-Vernon influence functional~\cite{FeynmanVernon}; we call this an influence matrix (IM), following Ref.~\cite{lerose2020}.

To write an explicit expression for the IM, we first introduce the {system-environment
decomposition of the Floquet operator}
$
\label{eq_SEint}
F = F_{\text{int}} F_S F_E
$
defined by the blue, red and grey gates in Fig.~\ref{fig_IM}b, respectively.
Thus, $F_S, F_E$ acts on the subsystem and environment only, such that $[F_S,F_E]=0$, while $F_{\rm int}$ is the interaction between them. Focusing for concreteness on the right environment $j\ge1$,
 the interaction Floquet operator is
$
F_{\text{int}} \equiv
V_{0,1}
$.
We define the partial  matrix elements of $F_{\text{int}}$ as the operators
$
[F_{\text{int}}]_{s,\sigma} \equiv
\braket{s | V_{0,1} | \sigma},
$
acting on the environment only,
conditioned on the input and output states of the subsystem  $\sigma$ and $s$, respectively.
With these notations, the influence matrix is defined as a discrete functional or tensor:
\begin{widetext}
\begin{equation}
\label{eq_IMdef}
\mathscr{I}[\sigma^\pm_\tau,s^\pm_\tau]
=
\Tr_E
\bigg(
[F_{\text{int}}]_{s^+_t,\sigma^+_{t}}
F_E
\cdots
F_E
[F_{\text{int}}]_{s^+_1,\sigma^+_1} F_E \; \rho_0^{E} \; F_E^\dagger
[F_{\text{int}}^\dagger]_{\sigma^-_1,s^-_1}
F_E^\dagger
\cdots
F_E^\dagger
[F_{\text{int}}^\dagger]_{\sigma^-_{t},s^-_t}
\bigg) \, ,
\end{equation}
\end{widetext}
where $\Tr_E \equiv \Tr_{j=1,2,\dots}$ and $\rho_0^{E} \equiv  \bigotimes_{j=1,2,\dots} \rho_0^{(j)}$.
{This object, and its left environment analog, are graphically highlighted by blue shading in Fig.~\ref{fig_IM}b}. We refer to components with $+,-$ as the ``forward'' and ``backward'' in time trajectory, respectively.

The influence matrix $\mathscr{I}$ contains full information about the dynamical effect of the environment on the subsystem.
In fact, arbitrary temporal correlations -- or outcomes of sequential measurements -- of observables involving the \textit{local} subsystem only (i.e., the qubit at $j=0$ here), can be expressed in terms of the subsystem internal dynamics $K_{j=0}$ and the IMs of the left and right environments.

The IM of a longer chain can be computed from that of a shorter one using the dual transfer matrix approach~\cite{Banuls09,lerose2020}: in Fig.~\ref{fig_IM}b the IM for the qubit $j=0$ is given by the IM for the qubit $j=2$, contracted with an extra vertical layer of gates above $j=1,2$ (the latter defines the dual transfer-matrix $\widetilde{T}$). Thus, the IM can be obtained by iteratively applying the dual transfer matrix, starting from the right boundary of the system.
For long chains with translationally invariant gate structure and initial state, the IM can thus be identified by a self-consistency equation, which takes the form of an eigenvector equation $\widetilde{T}\ket{\mathscr{I}}=\ket{\mathscr{I}}$.

To compute the IM, it is convenient to use its { interaction picture representation}, where the ``free'' environment evolution is absorbed to dress the interaction Floquet operators.
Specifically, we define
$
\widetilde{F}_{\text{int}}(\tau)
=
(F_E^\dagger)^\tau F_{\text{int}}
(F_E)^\tau
$.
Due to the strict light cone in local circuits, this is an operator with a finite support on qubits $0 \leq j \leq 2\tau$.
We can then rewrite Eq.~\eqref{eq_IMdef} as follows,
\begin{widetext}
\begin{equation}
\label{eq_IMintpict}
\mathscr{I}[\sigma^\pm_\tau,s^\pm_\tau]
=
\Tr_E
\bigg(
[\widetilde{F}_{\text{int}}(t)]_{s^+_{t},\sigma^+_{t}}
\cdots
[\widetilde{F}_{\text{int}}(1)]_{s^+_1,\sigma^+_1} \; \rho_0^E \;
[\widetilde{F}_{\text{int}}^\dagger(1)]_{\sigma^-_1,s^-_1}
\cdots
[\widetilde{F}_{\text{int}}^\dagger(t)]_{\sigma^-_{t},s^-_t}
\bigg)
\end{equation}
\end{widetext}
Due to the absence of correlations in $\rho_0^E$,  
the trace is taken effectively over a finite 
Hilbert space.

\section{Exact influence matrix of the kicked transverse-field Ising chain}\label{sec3}

In this Section, we derive an exact expression for the influence matrix of a family of integrable circuits, which belongs to the set of models defined by Eq.~\eqref{eq_circuit}. This family is defined by choosing Ising interactions and purely transverse single-qubit rotations. We work with the parametrization
\begin{equation}
\label{eq_KIC}
V_{j,j+1}= \exp \left(
i J Z_j Z_{j+1}
\right),
\qquad K_j = \exp \left(
i g X_j
\right),
\end{equation}
where $X_j,Y_j,Z_j$ denote Pauli matrices acting on qubit $j$;
the gates of this model are illustrated in Fig.~\ref{fig_IM}c.
The symmetries of the problem allow us to restrict the analysis to the quadrant $0\le J,g\le \pi/2$.
This paradigmatic Floquet model has been extensively investigated in several contexts~\cite{ProsenPRE02,Kim_ETH,Guhr1,BertiniSFF}.
The model can be solved by mapping the generators of the unitary gates to bilinear forms of fermionic creation/annihilation operators $c_j^\dagger$, $c_j$, via a Jordan-Wigner transformation.
This reduces it to a Floquet generalization of the Kitaev chain~\cite{Kitaev_2001}, i.e.,
\begin{equation}
\label{eq_kkc}
F=
\prod_{j\in\Z}
e^{
i J
(c^\dagger_j+c_j) (c_{j+1} -
 c^\dagger_{j+1})
}
\prod_{j\in\Z}
e^{
i g (c_j c^\dagger_j-c^\dagger_jc_j)
}
\, .
\end{equation}


The quasienergy spectrum $\pm \phi_k$ of this fermionic model as a function of the momentum $k$ is given by the following relation (see Appendix~\ref{app_diagonalizationKIC} for details):
\begin{equation}\label{eq:phi_k}
\cos \phi_k=\cos 2J \cos 2g + \sin 2J \sin 2g \cos k .
\end{equation}
The two quasienergy bands are generally separated by a gap, which closes at $k=0$ or $k=\pi$ when $J=g$ or $J=\pi/2-g$. The gap closing signals a phase transition between distinct topological Floquet phases, some of which feature edge modes with $\phi^e=0,\pi$ (arising for $g < J$ and $g > \pi/2- J$, respectively~\cite{DuttaPRB13}). This, as we will show below, has an imprint on the structure of the IM. Interestingly, at the intersection between these two critical lines, $J=g=\pi/4$ (the self-dual point), the quasienergy spectrum becomes linear everywhere in the Brillouin zone, $\phi_k = k$~\cite{Guhr1,BertiniSFF}, signaling equivalence of space and time propagation. 

Turning to the computation of the IM, we first rewrite the environment Floquet operator $F_E$ in terms of the creation/annihilation operators $d^\dagger_m$, $d_m$ of quasiparticle modes with quasienergy $\phi_m$. These are the eigenmodes of the kicked Kitaev model in Eq.~\eqref{eq_kkc} defined on a half-chain with open boundary conditions, $j\geq 1$, and are related to the original fermionic operators  $c_j^\dagger$, $c_j$ by a Bogoliubov-de Gennes transformation. The index $m$ collects both the continuous momentum $k$ and the possible discrete edge modes $e$. This representation allows us to express the interaction-picture evolution of the subsystem-environment interaction operator in Eq.~\eqref{eq_IMintpict} as follows,
\begin{equation}
\label{eq_fermionicintpict}
 \widetilde{F}_{\text{int}}(\tau)
=
\exp \bigg[  J
\big(c_0+c_0^\dagger\big)
\bigg(
\sum_{m}
\mathcal{C}_{m}
e^{-i \tau \phi_m } \; d_m
+ \text{H.c.}
\bigg)
\bigg] ,
\end{equation}
where
$
\{\mathcal{C}_{m}\}
$
are the coefficient  of $\{d_{m}\}$ in the expansion of the boundary operator $i(c_1-c_1^\dagger)$.
Their explicit form 
can be found in Appendix A.

The influence matrix in Eq.~\eqref{eq_IMintpict} with the interaction-picture operators in Eq. \eqref{eq_fermionicintpict} becomes a trace over the fermionic Fock space spanned by the environment modes $\{ d_m,d^\dagger_m \}$, parametrically depending on the configuration of the subsystem fermion $c_0,c^\dagger_0$ at all $\tau=0,...,t$.  This trace can be computed using its representation as a multiple convolution of Gaussian Grassmann kernels~\cite{ItzyksonDrouffe}, obtained by inserting resolutions of the identity by fermionic coherent states at each $\tau$.
%
%
%
The resulting Grassmann influence functional can be viewed as a ``many-body
wavefunction'' in the fermionic Fock space spanned by the tensor product of
all input and output subsystem Hilbert spaces along the closed-time Schwinger-Keldysh
contour.
This temporal Fock space is generated by four creation and four annihilation fermionic operators $f^{(\dagger)}_{\uparrow/\downarrow,+/-}$, characterized by two ``flavors'' per temporal lattice site $\tau$, input-output and forward-backward [see
Fig.~\ref{fig_IM}(b)], labelled by subscripts $\uparrow,\downarrow$ and $+,-$, respectively.
%

Focusing on infinite-temperature initial ensembles $\rho_0^{(j)}=\mathbb{1}/2$~\cite{lerose2020},
%
and evaluating the Grassmann path integral,
we obtain a compact formula for the exact IM (see Appendix~\ref{app_derivationIM} for the derivation).
%
The resulting IM wavefunction in the second-quantized language is obtained by substituting the Grassmann variables 
by the corresponding creation operators $f^\dagger_{\uparrow+,\tau},{f}^\dagger_{\downarrow+,\tau},f^\dagger_{\uparrow-,\tau},{f}^\dagger_{\downarrow-,\tau}$, which yields:  
\begin{widetext}
\begin{equation}
\label{eq_IMoperator}
\ket{\mathscr{I}} \; \propto \;
e^{
\sum_\tau
\big(
f^\dagger_{\uparrow+,\tau} {f}^\dagger_{\downarrow+,\tau}
-
f^\dagger_{\uparrow-,\tau} {f}^\dagger_{\downarrow-,\tau}
\big)
+ 
\sum_{\tau,\tau'}
\kappa(\tau'-\tau)
\big[
f^\dagger_{\uparrow+,\tau} f^\dagger_{\uparrow-,\tau'}
+
\Theta(\tau'-\tau)
\big( f^\dagger_{\uparrow+,\tau} f^\dagger_{\uparrow+,\tau'}
-   f^\dagger_{\uparrow-,\tau} f^\dagger_{\uparrow-,\tau'} \big)
\big]
}
\ket{\emptyset}  .
\end{equation}
Here, $\ket{\emptyset}$ is the fermionic vacuum state, $\tau,\tau'=1,\dots,t$ label the temporal lattice sites, and $\Theta(\tau)=[1+\sign(\tau)]/2$ is Heaviside's theta function. 
Finally, the real function, which fully encodes the effect of the environment,
\begin{equation}
\label{eq_gamma}
    \kappa(\tau) = 2 (\tan J)^2
    \bigg[
\int_{0}^{\pi} \frac{dk}{2\pi}
\big\lvert \mathcal{C}_{k}  \big\rvert^2 \cos(\phi_k \tau) + |\mathcal{C}_{e=0}|^2
+ |\mathcal{C}_{e=\pi} |^2 (-)^\tau \bigg]
\equiv \int_{-\pi}^\pi \frac{d\omega}{2\pi} \mathcal{J}^R(\omega) e^{i\omega \tau}
\end{equation}
\end{widetext}
characterizes the ``correlations'' between temporally separated subsystem's configurations both on the same and on the opposite branch of the Keldysh contour. In Eq.~(\ref{eq_gamma}) we have introduced a function $\mathcal{J}^R(\omega)$, which is the fermionic analog of the environment's spectral density, discussed in the context of open quantum systems for a bath of harmonic oscillators~\cite{LeggettRMP}. Furthermore, $\mathcal{C}_{e=0,\pi}$ arise due to the edge modes with $\phi^e=0,\pi$.

For the homogeneous kicked Ising chain considered here, the spectral density has a continuous part supported in the positive and negative quasienergy band $\omega \in \pm [ 2|J - g|, 2|J + g| ]$ (for $J+g<\pi/2$) or $\pm[ 2|J - g|, 2\pi-2|J + g| ]$  (for $J+g>\pi/2$).
The coefficients $\mathcal{C}_k$ vanish at the quasienergy band maxima/minima at $k=0,\pi$, which, in combination with the van Hove singularities in the density of states, gives rise to $\mathcal{J}^R(\omega)$ vanishing as a square root at the band edges $\omega^*$, $\mathcal{J}^R(\omega)\propto \sqrt{|\omega-\omega_*|}$.
In the topologically non-trivial phases, the edge modes produce additional contributions proportional to $|\mathcal{C}_{e=0}|^2 \delta(\omega)$ and/or $|\mathcal{C}_{e=\pi}|^2 \delta(\omega-\pi)$ in $\mathcal{J}^R(\omega)$.

The non-local in time influence of the environment on the subsystem's dynamics  is encoded in the function $\kappa(\tau'-\tau)$.
Physically, $\kappa$ can be interpreted as a response function of the environment to a boundary perturbation.
 For generic parameter values in the topologically trivial phase $J<g<\pi/2-J$, at large time separations $\tau\to\infty$ this function displays oscillatory behavior at frequencies $\omega^*$ modulated by a power-law decay $\kappa(\tau) \sim (\tau/\tau_0)^{-3/2}$. This behavior originates from the square-root form of $\mathcal{J}^R(\omega)$ near the band edges.
These slowly decaying correlations between temporally separated subsystem configurations can be thought of as mediated by environment excitations with vanishing velocity ($\partial_k \phi|_{k=0}=\partial_k \phi|_{k=\pi}=0$), residing in the vicinity of band edges.


At criticality, $J=g$ ($J=\pi/2-g$), the quasienergy spectrum undergoes a transformation, which modifies the continuous part of the spectral density compared to the generic case. Specifically, the quasienergy gap closes at $k=0$ ($k=\pi$), and quasiparticles can travel at a finite speed $|\partial_k \phi| \sim c >0$ down to $k=0$ ($k=\pi$).
The gap closing leads to suppression of the corresponding power-law contribution in the spectral density and in the function $\kappa(\tau)$.

Strikingly, at the doubly-critical self-dual point $J=g=\pi/4$, both $k=0$ and $k=\pi$ band edges disappear, as the spectrum becomes linear throughout the Brillouin zone. The environment's influence thus becomes local in time  $\kappa(\tau) = 2 \delta_{\tau,0}$, which underlies the perfect dephaser property of the system, corresponding to an exactly Markovian (i.e., memoryless) dynamics of subsystems interacting with the environment~\cite{lerose2020}. 
Detuning from such special point, the relaxation dynamics of subsystems acquires a finite memory time.

In the topologically non-trivial phases the edge modes are associated with discrete points in the quasienergy spectrum, giving rise to additional \textit{undamped} contributions to $\kappa(\tau)$ with frequency $0$ or $\pi$, cf. Eq.~\eqref{eq_gamma}.
These long-range temporal correlations express the memory of the initial condition at the boundary of an open chain, due to conserved operators exponentially localized near the edge. In the fermionic representation, these are Majorana edge modes~\cite{Kitaev_2001,DuttaPRB13}. In the original spin degrees of freedom, they correspond to strong zero modes~\cite{FendleyLongCoherenceEdge}.
Remarkably, structural information on non-trivial \textit{edge} physics shows up in the \textit{bulk} IM. In Sec.~\ref{sec5} below we will show that this change in the IM across a phase transition can be characterized by temporal entanglement.

The behavior of the function $\kappa(\tau)$ in the four distinct phases is illustrated in the four insets of Fig.~\ref{fig_arealaw}.


 \section{Area-law temporal entanglement}\label{sec4} 

\begin{figure*}
    \centering
    \includegraphics[width=0.7\textwidth]{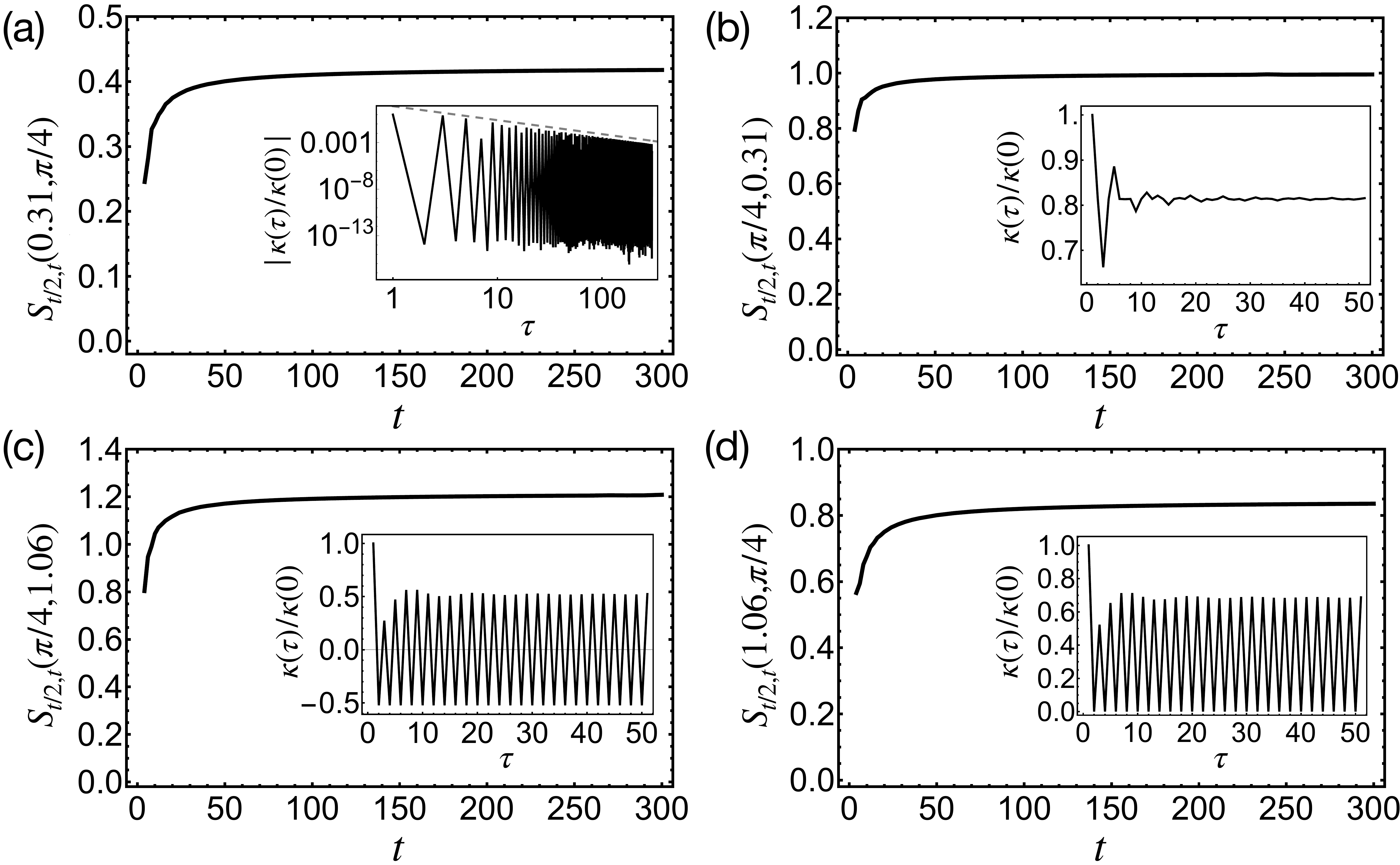}
    \caption{
    Area-law scaling of temporal entanglement entropy illustrated in different regions of the phase diagram.
    The four panels (a-d) show the scaling of the maximum (half-chain) bipartite von Neumann entropy of the IM as a function of evolution time for the four choices of parameter values marked by black stars in Fig.~\ref{fig_integrable}a, belonging to the four topologically distinct Floquet phases. In all cases, long-time saturation is evident, indicating the area law.
    Insets show the behavior of the function $\kappa(t)$ in the four cases, which illustrates the discussion in the main text. [Here we rescaled $\kappa(\tau)$ by its prefactor $\kappa(0)\equiv 2 \tan^2 J$ in Eq.~\eqref{eq_gamma}.] In panel (a), the dashed straight line in log-log scale represents the power-law decay, $\tau^{-3/2}$; in the other panels, in addition to this power-law decaying contribution, $\kappa(\tau)$ includes a constant (b,d) and/or alternating (c,d) term, due to the $0$ and/or $\pi$ edge modes, respectively.}
    \label{fig_arealaw}
\end{figure*}

 In this Section, we will analyze the temporal entanglement properties of the IM
 ``wavefunction'' computed above, Eq.~(\ref{eq_IMoperator}), which has a Gaussian
 form with power-law decaying correlations.
 We will be interested in its bipartite von Neumann entanglement entropy, and its scaling with
 the evolution time.\footnote{It is important to note that unlike
 regular wavefunctions, the IM normalization is such that the Keldysh ``partition function''
 (the path integral without observables) is unity. In this paper, however, to compute its von Neumann entropy we normalize the IM
 as a regular wavefunction, which involves rescaling it by a factor exponentially large in $t$.}
 By computing it numerically up to long times $t$, we will
 demonstrate that TE entropy remains bounded as $t\to\infty$, and thus it
 obeys area-law scaling. We will further support this conclusion by an analytic
 argument demonstrating that the IM wavefunction can be viewed as the ground state
 of a gapped quadratic Hamiltonian with algebraically decaying couplings; such states have
 been rigorously proven to exhibit area-law entanglement entropy~\cite{Korepin09Review,Its08CMP} (see below).

Viewed as a quantum state of a fermionic chain, the IM wavefunction in Eq.~(\ref{eq_IMoperator}) is a pure Gaussian state.
The entanglement entropy associated with a bipartition between a subset $A$ of the lattice (of size $|A| \le t/2$) and its complement is $S_{A}=-\Tr(\rho_A \log \rho_A)$, where $\rho_A$ is the reduced density matrix of subsystem $A$. For Gaussian states, $\rho_A$ is uniquely determined by the two-body correlations within $A$, compactly collected in the hermitian matrix
$C_{A}\equiv\big(C_{i,j}\big)_{i,j \in A}$, with~\cite{Latorre_2009}
\begin{equation}
C_{i,j} =
\begin{pmatrix}
\frac{\langle \mathscr{I} | f_i f^\dagger_j | \mathscr{I}\rangle}{\langle\mathscr{I}|\mathscr{I}\rangle} &
\frac{\langle \mathscr{I} | f_i f_j | \mathscr{I}\rangle}{\langle\mathscr{I}|\mathscr{I}\rangle} \\
\frac{\langle \mathscr{I} | f^\dagger_i f^\dagger_j | \mathscr{I}\rangle}{\langle\mathscr{I}|\mathscr{I}\rangle} &
\frac{\langle \mathscr{I} | f^\dagger_i f_j | \mathscr{I} \rangle}{\langle\mathscr{I}|\mathscr{I}\rangle} \\
\end{pmatrix}.
\end{equation}
Here indices $i,j$ range in $(a,b,\tau)$ with $a=\uparrow,\downarrow$, $b=+,-$, and $\tau \in A$ (with a slight abuse of notation). Considering a half-chain bipartition corresponds to choosing ${1\le \tau \le t/2}$.
Entropy is computed
as $S_{|A|,t} = - \sum_{i=1}^{4|A|} p_i \log p_i + (1-p_i) \log (1-p_i)  $, 
where the binary probability $(p_i,1-p_i)$ associated with the $i$-th pair of eigenvalues of $C_{A}$  represents the uncertainty in the occupation of the half-chain single-particle orbital defined by the corresponding pair of eigenvectors.

\begin{figure*}
    \centering
    \includegraphics[width=\textwidth]{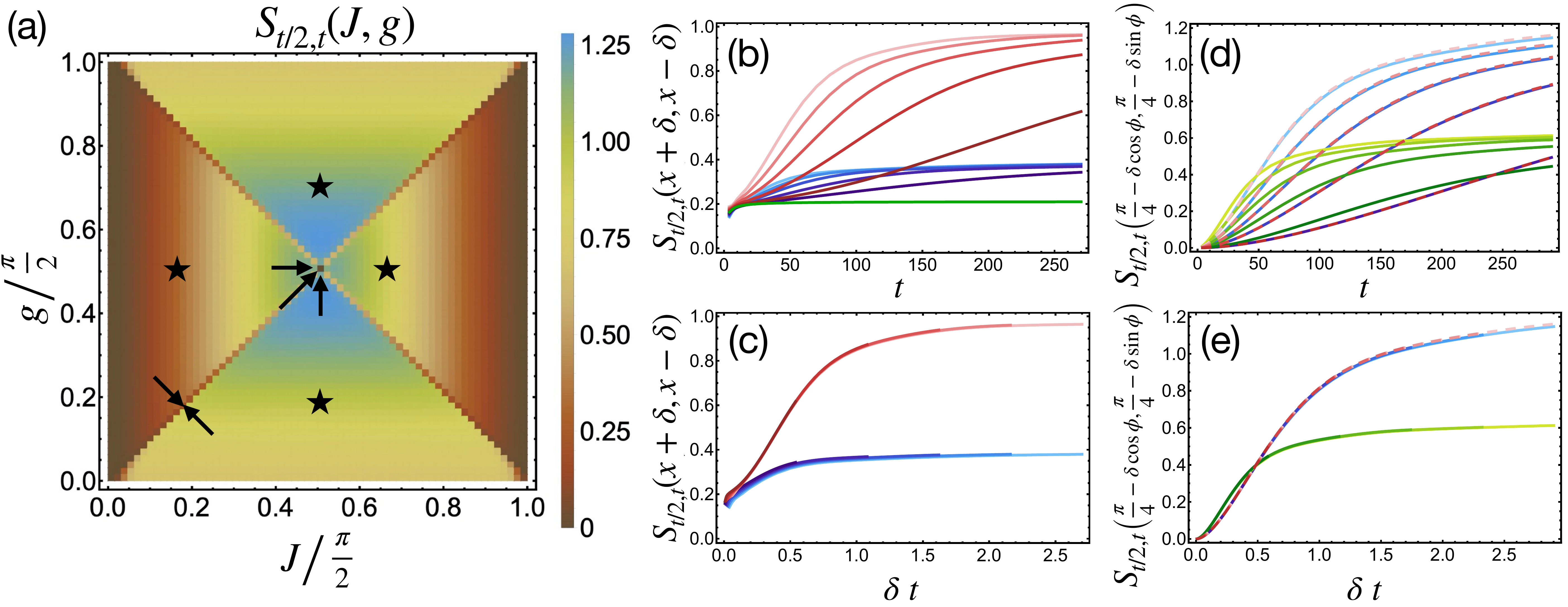}
    \caption{
    Scaling behavior of temporal entanglement in the transverse-field kicked Ising chain.
    Panel (a): Saturated value of the half-chain bipartite TE entropy across the parameter space (here we have fixed $t=150$). Discontinuities appear at the critical lines $J=g$, $J=\pi/2-g$.
    Panel (b-e): scaling behavior upon approaching the critical lines (b,c) and the self-dual point (d,e), as indicated by the black arrows in panel (a).
    Panels (b,c) describe the vicinity of the point $(J,g)=(x,x)$ with $x=0.31$. Detunings are $\delta=0.01\div0.002$ (lightest to darkest red tones), $\delta=-0.01\div-0.002$ (lightest to darkest blue tones), $\delta=0$ (green).
    Panel (c) shows the collapse of the curves upon linearly rescaling time by the detuning $\delta$ from the critical point.
    Panels (d,e) describe the vicinity of the self-dual point $(J,g)=(\pi/4,\pi/4)$. Detunings are $\delta=0.01\div0.002$
    in the direction $\phi=0$
    (lightest to darkest blue tones, solid lines), $\phi=\pi/4$
    (lightest to darkest green tones, solid lines),
    $\phi=\pi/2$
    (lightest to darkest red tones, dashed lines).
    Panel (e) shows the collapse of the curves upon linearly rescaling time by the detuning $\delta$ from the self-dual point.
    }
    \label{fig_integrable}
\end{figure*}

As one of the central results of this Article, we find that the maximum TE entropy $S_{t/2,t}(J,g)$ saturates to a finite value $S_{\infty}(J,g)$ as $t\to\infty$ for all parameters values $0\le J,g \le \pi/2$.
Several instances of this area-law scaling are reported in Fig.~\ref{fig_arealaw}.
The saturation value of the TE entropy as a function of $J,g$ is illustrated in Fig.~\ref{fig_integrable}a, where points marked by black stars indicate the parameter choice of Fig.~\ref{fig_arealaw}. The pattern of saturated TE mimics the phase diagram of the model, exhibiting jumps across critical lines; in the following Section we will elucidate the origin of this behavior, and analyze the scaling of TE near critical points.

We have verified that the saturation value of TE entropy is independent of the precise position of the bipartition cut, which suggests that the IM wavefunction can be expressed as the ground state of a gapped quasilocal Hamiltonian $\widetilde{\mathcal{H}}$.
Here we construct such a quadratic parent Hamiltonian, which can be viewed a self-adjoint deformation of the non-Hermitian generator of the dual transfer matrix $\widetilde{T}$. For simplicity, we first focus on the topologically trivial phase $J<g<\pi/2-J$. We take the limit $t\to\infty$ and represent the state in Eq.~(\ref{eq_IMoperator}) with Keldysh-rotated fields $f_{\uparrow[\downarrow]cl,\tau} = \frac 1 {\sqrt{2}} (f_{\uparrow[\downarrow]+,\tau} + f_{\uparrow[\downarrow]-,\tau})$, $f_{\uparrow[\downarrow]q,\tau} = \frac 1 {\sqrt{2}} (f_{\uparrow[\downarrow]+,\tau} - f_{\uparrow[\downarrow]-,\tau})$ \cite{Kamenev} (see also Appendix~\ref{app_derivationIM}) and in the frequency domain, neglecting the boundary effects at $\tau=0,t$:
\begin{widetext}
\begin{equation}
    \ket{\mathcal{\mathscr{I}}}
    =
    \exp\bigg\{
    \int_{-\pi}^\pi
    \frac{d\omega}{2\pi}
    \Big[
    f^\dagger_{\uparrow,cl}(\omega)
    f^\dagger_{\downarrow,q}(-\omega)
    +
    f^\dagger_{\uparrow,q}(\omega)
    f^\dagger_{\downarrow,cl}(-\omega)
    +
    \mathcal{J}(\omega) \;
    f^\dagger_{\uparrow,q}(\omega)
    f^\dagger_{\uparrow,cl}(-\omega)
    \Big]
    \bigg\} \ket{\emptyset} \, .
    \label{eq_BCSform}
\end{equation}
\end{widetext}
Here $\mathcal{J}(\omega)=\mathcal{J}^R(\omega)+ i \mathcal{J}^I(\omega) $ is the Fourier transform of $\Theta(\tau)\kappa(\tau)$ (NB here $\Theta(0)\equiv 1/2$). The imaginary part  $\mathcal{J}^I(\omega)$ is related to the spectral density  $\mathcal{J}^R(\omega)$ in Eq.~(\ref{eq_gamma}): $\mathcal{J}^I(\omega)=\mathcal{P}\int_{-\pi}^\pi \frac{d\Omega}{2\pi} \cot[(\Omega-\omega)/2] \mathcal{J}^R(\Omega)$ (principal part prescription). Thus, $\mathcal{J}(\omega)$ is a continuous function (up to the effect of the edge modes which is discussed below), with square-root singularities at the quasienergy band edges.

The $8\times8$ antisymmetric block $(\omega,-\omega)$ in the exponent of Eq.~\eqref{eq_BCSform} can be made real by eliminating the complex phase $\varphi(\omega)$ of $\mathcal{J}(\omega)=\rho(\omega)e^{i\varphi(\omega)}$ via, e.g. a suitable redefinition of the phases of $f^\dagger_{\uparrow,q}(\omega)$ and $f^\dagger_{\downarrow,cl}(-\omega)$. The resulting real antisymmetric quadratic form can be diagonalized by an orthogonal transformation $R(\omega)$, which brings it to the Bardeen-Cooper-Schrieffer-like pairing form $i \sum_{\alpha=1}^4 \tan(\theta_{\alpha})  b^\dagger_{\alpha}(\omega)  b^\dagger_{\alpha}(-\omega)$.
 Thus, the Bogolubov modes
 $ \tilde b_{\alpha}(\pm\omega)= \cos(\theta_{\alpha}) b_{\alpha}(\pm\omega)
 + i \sin(\theta_{\alpha})
b^\dagger_{\alpha}(\mp\omega)$ annihilate the state $\ket{\mathscr{I}}$.
 We can use them to construct gapped quadratic parent Hamiltonians for $\ket{\mathscr{I}}$ as
 \begin{equation}
 \label{eq_parentH}
     \widetilde{\mathcal{H}} = \sum_
 {\alpha=1}^4 \int_{-\pi}^\pi \frac{d\omega}{2\pi}
 \tilde\epsilon_\alpha(\omega) \,
  \tilde b^\dagger_{\alpha}(\omega) \tilde b_{\alpha}(\omega),
 \end{equation}
 where $\tilde\epsilon_\alpha(\omega)>0$ are arbitrary positive functions, e.g., $\tilde\epsilon_\alpha(\omega)\equiv 1$.

The expression of $\widetilde{\mathcal{H}}$ in terms of the original lattice degrees of freedom $f_{\uparrow/\downarrow,+/-,\tau},f_{\uparrow/\downarrow,+/-,\tau}^{\dagger}$ can be obtained from Eq.~\eqref{eq_parentH}  by an inverse rotation of operators $\tilde b_{\alpha}(\pm\omega)$, followed by Fourier transformation. The inverse rotation involves the Bogolubov angles $\theta_\alpha(\omega)$, the rotation $R(\omega)$, and the phase factor $e^{i\varphi(\omega)}$.
By the standard properties of the Fourier transform, the degree of locality of $\widetilde{\mathcal{H}}$ on the temporal lattice is determined by the degree of smoothness of those quantities in the frequency domain. We note that the quantities $\theta_\alpha(\omega)$ and $R(\omega)$ arise from the diagonalization of a real antisymmetric matrix that depends analytically on $\rho(\omega)$, hence they can be chosen to depend smoothly and periodically on $\rho(\omega)$ for $\omega \in [-\pi,\pi]$.
Furthermore, as shown above, $\rho(\omega)$ and $\varphi(\omega)$ themselves have square-root singularities at the band edges, and smoothly depend  on $\omega$ elsewhere.
Thus, the Fourier transform of all terms in $\widetilde{\mathcal{H}}$ is guaranteed to decay algebraically as $ |\tau-\tau'|^{-3/2}$~\footnote{A singularity $|\omega-\omega^*|^\alpha$ gives rise to an asymptotic contribution $|\tau|^{-1-\alpha}$ to the Fourier transform at large $|\tau|$.}.

 The above procedure has allowed us to construct a family of gapped, quasilocal quadratic parent Hamiltonians [one for each choice of  positive smooth periodic functions $\epsilon_\alpha(\omega)$ in Eq.~\eqref{eq_parentH}].
 The absence of discontinuous jumps in the frequency domain gives rise to an area-law scaling of temporal entanglement without logarithmic corrections, as implied by the so-called Widom theorem on the asymptotic behavior of block Toeplitz determinants~\cite{Korepin09Review,Its08CMP}. Area-law scaling has been previously found numerically~\cite{VodolaPRL14} for non-critical Kitaev chains with couplings algebraically decaying with the distance $r$ as $r^{1+\sigma}$, $\sigma>0$. We envisage that the saturation value $S_{A,t}(J,g)$ discussed here may be computed analytically by extending the techniques of Refs.~\cite{Korepin09Review,Its08CMP,AresPRA18}.
 We further note that a proof of area-law entropy scaling for the ground state of general (non-quadratic) one-dimensional gapped Hamiltonians with algebraically  decaying couplings $r^{1+\sigma}$, $0<\sigma<1$, lies beyond currently available rigorous results~\cite{Kuwahara20AreaLawLR}.

 In the topologically non-trivial phases, the presence of edge modes produces additional delta-function contributions in $\mathcal{J}(\omega)$ at $\omega=0$ and/or $\pi$. Physically, the influence of an edge mode can be understood as arising from coupling the subsystem to an additional isolated particle. The effect of such a coupling on the influence matrix is expressed by the action of an infinite-range operator $\exp[4 (\tan J)^2|\mathcal{C}_{0,\pi}|^2 \sum_{\tau<s} (\pm)^{\tau-\tau'} f^\dagger_{\uparrow,q,\tau}f^\dagger_{\uparrow,cl,\tau'}]$, which can be factored out in Eq.~\eqref{eq_IMoperator}. This operator has finite Schmidt rank relative to the half-chain bipartition,
 and thus cannot generate violations of the area-law scaling of TE entropy.
 It does, however, introduce \textit{long-range} temporal correlations and entanglement.

\section{Temporal entanglement scaling near critical points}
\label{sec5} 

The saturated value of TE entropy, illustrated in Fig.~\ref{fig_integrable}, mimics the phase diagram of the model. In particular, there is a finite jump of  $S_\infty$ across the phase boundaries between the Floquet topological phases. At the first glance, this behavior seems surprising, as the influence matrix is a characteristic of the infinite-temperature dynamics of the system, whereas the singular scaling of spatial entanglement, which is typically used to detect quantum phase transitions, concerns the low-energy sector only.

These TE jumps can be attributed to the singular changes in the spectral density $\mathcal{J}^R(\omega)$ when crossing a critical point: indeed, as discussed above, the band edges are responsible for the power-law decaying interactions in the influence matrix. A gap closing at the critical point leads to the disappearance of a band edge, which modifies the spectral density, giving rise to a different value of saturated TE entropy. Approaching a critical point $(J,g)=(x,x)$ or $(x,\pi/2-x)$ from two opposite sides, $S_{t/2,t}$ saturates at two distinct saturation values as $t\to\infty$.


Further, we note that $S_{t/2,t}$ depends continuously on the parameters $J,g$ for a fixed $t$, and therefore the convergence to $S_\infty$ as $t\to\infty$ must become increasingly slow as the critical line is approached, suggesting the onset of scaling behavior. We investigate this by computing $S_{t/2,t}(x-\delta,x+\delta)$ for a sequence of positive, vanishing and negative detunings $\delta$, and for a range of time windows $t$.
[We have verified similar behavior for the other critical line $(J,g)=(x,\pi/2-x)$.]
The collapsed plots in Fig.~\ref{fig_integrable}b-c nicely confirm the scaling hypothesis, demonstrating that
\begin{equation}
    S_{t/2,t}(x-\delta,x+\delta) \underset{\substack{\delta\to0^\pm\\t\to\infty}}{\thicksim} F(x,\delta \cdot t)
\end{equation}
where the scaling function $F$ satisfies $F(x,0)=S_\infty(x,x)$, $F(x,\pm\infty)=S_\infty(x-0^\pm,x+0^\pm)$.

Finally, the doubly-critical self-dual point, $J=g=\pi/4$, being a perfect dephaser, has $S_{\tau,t}(\pi/4,\pi/4)\equiv0$. The region around it contains multiple scaling behaviors, depending on the direction $\phi\in[0,2\pi)$ of the detuning, i.e.,
\begin{equation}
    S_{t/2,t}(\pi/4-\delta\cos\phi,\pi/4-\delta\sin\phi) \underset{\substack{\delta\to0^+\\t\to\infty}}{\thicksim} G(\phi,\delta \cdot t).
\end{equation}
The scaling function satifies $G(\phi,0)=0$, $G(\phi,\infty)=S_\infty(\pi/4-\cos\phi \, 0^+,\pi/4-\sin\phi \,0^+)$.
We have found that $G(\phi,\infty)$ takes a constant value in the four quadrants $n\pi/4 < \phi < (n+1)\pi/4$, $n\in\mathbb{Z}$, and jumps discontinuously to a distinct value when $\phi$ is a multiple of $\pi/4$ --- see Fig.~\ref{fig_integrable}d-e.
This finding sheds light on the parametrically slow growth of TE entropy in models detuned from a perfect dephaser point, first reported in Ref.~\cite{lerose2020} (see also the next Section on integrability breaking).


\section{Integrability breaking}\label{sec6} 

\begin{figure*}
    \centering
    \includegraphics[width=0.98\textwidth]{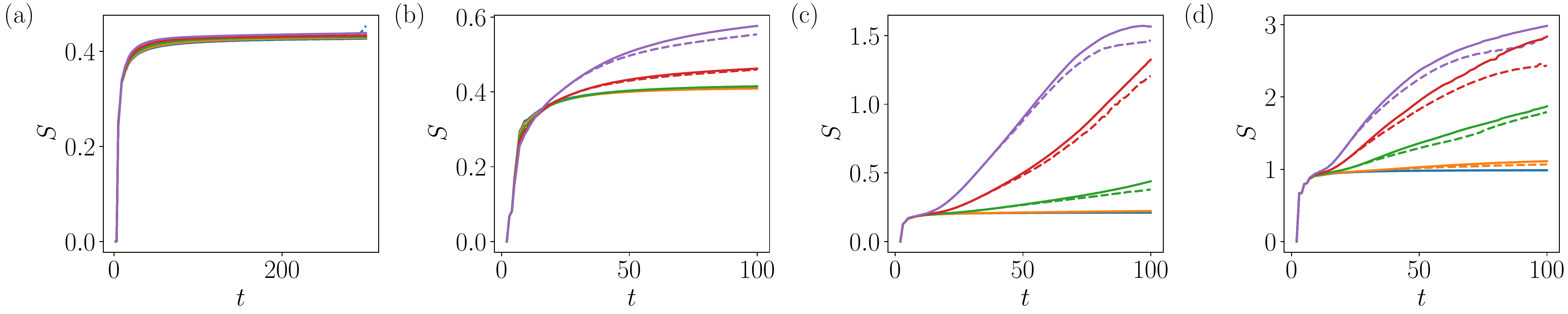}
    \caption{
        Scaling of TE entropy vs evolution time for various choices of parameters $J,g$ [(a):
        $J=0.31,g=\frac{\pi}{4}$; (b): $J=0.31,g=0.5$; (c): $J=0.31,g=0.31$; (d):
        $J=\frac{\pi}{4},g=0.31$; 
        parameter values of (a), (c), (d) are also marked in the diagram in Fig.~\ref{fig_integrable}a] and a range of integrability breaking perturbations
        $h=0.02,0.06,0.10,0.14,0.18$ (curves from bottom to top). The results for bond dimension $\chi=64$ (dashed) and $\chi=128$
        (solid) are well converged up
        to large numbers of Floquet driving periods. We observe that for low
        $h$, the curves are hardly distinguishable from the area-law behavior of the nearby integrable system (cf. the corresponding curves in Figs.~\ref{fig_arealaw}a,b, ~\ref{fig_integrable}b). Remarkably, in panel (a) this apparent saturation persists even for the strongest integrability breaking parameter we considered (note the longer time scale).
    }
    \label{fig_ent}
\end{figure*}

\begin{figure*}
    \centering
    \includegraphics[width=0.98\textwidth]{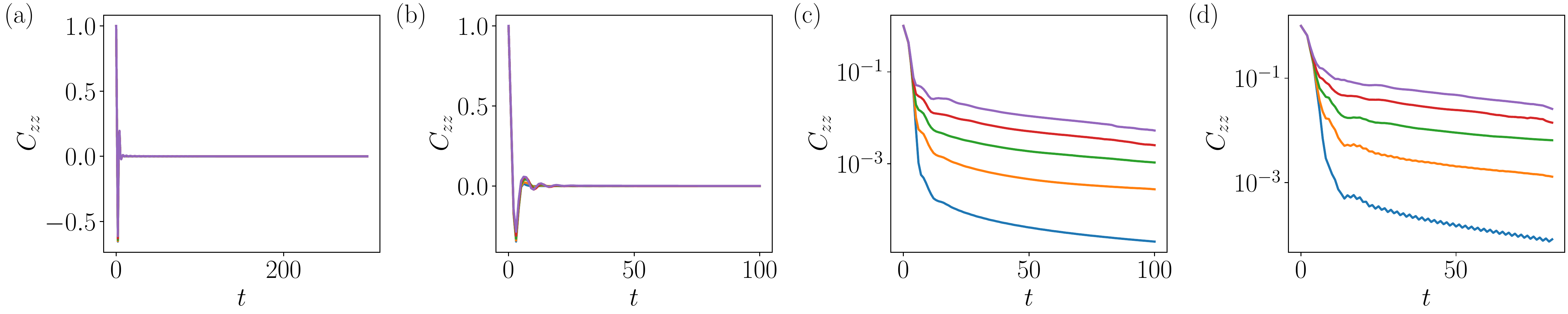}
    \caption{
        Autocorrelation function $C_{zz}(t)$ [cf. Eq.~\eqref{eq_czz}] for
        parameters $J=0.31,g=\frac{\pi}{4}$ (a), $J=0.31,g=0.5$ (b),
        $J=0.31,g=0.31$ (c) and $J=\frac{\pi}{4},g=0.31$ (d) as well as for
        increasing strength of integrability breaking perturbation
        $h=0.02,0.06,0.10,0.14,0.18$ (curves bottom to top). Parameter values are the
        same as in the corresponding panels of Fig.~\ref{fig_ent}. For panels (c) and (d), the autocorrelation
        drops rapidly until a slower decay takes over at a time that decreases with the strength of
        integrability breaking.
    }
    \label{fig_czz}
\end{figure*}

We are now in a position to address the effects of integrability breaking on the scaling of TE.
To preserve the simple structure of the circuit dynamics in Eq.~(\ref{eq_KIC}), we perturb the direction of the kick with a small
longitudinal field $h$, such that
\begin{equation}
    K_j \mapsto K'_j=\exp(igX_j)\exp(ihZ_j).
\end{equation}
This model is quantum-chaotic at generic parameter values~\cite{Kim_ETH}. The
perturbation operator $Z_j$ maps to a non-quadratic (and non-local) fermionic
operator within the Jordan-Wigner transformation, destroying integrability and
precluding a general analytical solution for the IM. We thus resort to numerical
computations.

Refs.~\cite{Banuls09,muller2012tensor,HastingsPRAFolding} pioneered the use of
MPS methods as a numerical tool to compute subsystem dynamics via transverse contraction of the tensor network, finding that
this approach is efficient in certain parameter regimes. Ref.~\cite{lerose2020}
and subsequently Refs.~\cite{Chan21,SonnerAoP} exploited a similar numerical approach
within the influence-functional formalism. Ref.~\cite{lerose2020}, in particular, used an MPS
ansatz to approximate the IM wavefunction in a neighborhood of the self-dual
point $J=g=\pi/4$. Remarkably, TE entropy vanishes exactly for \textit{arbitrary} $h$ at this special point, due to the perfect dephaser property of the system. In a neighborhood of this point TE entropy scales slowly with evolution time (cf. Fig.~\ref{fig_integrable}d),
which makes the MPS ansatz efficient. 
In Sec.~\ref{sec4} above, we showed that the TE entropy obeys an area-law scaling at integrability ($h=0$)
throughout the phase diagram. This suggests that the MPS ansatz also provides an
efficient representation of the self-consistent IM when the  integrability
breaking parameter $h$ is sufficiently small and the evolution time $t$ is short enough.

To understand how integrability breaking modifies the scaling of TE, we use the
MPS approach~\cite{lerose2020} to compute the IM for several values of $J,g$, and integrability breaking parameter $h$. To perform the computation, we represent the dual
transfer matrix as an MPO (of bond dimension $4$) and iteratively applying it to a boundary IM in an MPS form, compressing it to a fixed maximum bond dimension $\chi$ after each iteration. Our code makes use of the tenpy library~\cite{tenpy}. After at most $t$ iterations the thermodynamic
IM is reached due to the strict light cone effect in this model.
To avoid intermediate states of high entanglement encountered during iterations~\cite{Chan21,SonnerAoP} we
choose a perfect dephaser~\cite{lerose2020} boundary IM, which makes the approach significantly more efficient~\cite{SonnerAoP}.

A selection of numerical results is reported in Fig.~\ref{fig_ent}. Convergence
with respect to increasing the bond dimension is shown by comparing the results
for $\chi=64$ (dashed lines) and $128$ (solid lines). This shows that the method
produces reliable results for subsystems' dynamics over a very large number of
Floquet cycles, from several tens to few hundreds depending on the parameter
values, despite the breaking of integrability. {We remark that the
strength of integrability breaking  considered here is strong enough to show signatures 
of quantum ergodicity and chaos according to conventional probes
such as level spacing statistics for system sizes as low as $L\approx 12$.}

The behavior of TE scaling of the non-integrable model shows a visible 
dependence on the phase diagram of the nearby integrable limit. Results in
Fig.~\ref{fig_ent}a,b) concern two points inside the paramagnetic phase
$J<g<\pi/2-J$. Our numerical results indicate that the scaling of TE of this
non-integrable model is compatible with a long-time saturation. For
$g=\frac{\pi}{4}$, TE entropy is almost insensitive to $h$ even for the
strongest integrability breaking strength considered ($h=0.2$), comparable to
the magnitude of $J=0.31$.

Moving towards the symmetry-broken phase, the behavior of TE entropy 
becomes sensitive to $h$, as shown in Fig.~\ref{fig_ent}b and further in Fig.~\ref{fig_ent}c
(critical line $J=g$, marked by black arrows in Fig.~\ref{fig_integrable}a) and
in Fig.~\ref{fig_ent}d (deep inside the symmetry-broken phase $g<J<\pi/2-g$, bottommost point
marked by a black star in Fig.~\ref{fig_integrable}a).
A clear slow growth above the area-law saturation value of the $h=0$ limit appears when $h$ is increased.
For small $h$, we find that TE entropy first converges to the saturation value
of the model with $h=0$, subsequently slowly increasing at a
rate that grows as a function of $h$.

These results indicate the possibility to efficiently simulate the transient local relaxation dynamics
of systems close to integrability throughout the phase diagram.
We note that our present data do not allow to conclude whether the growth of $S$ (when present) persists as $t\to\infty$.
We speculate that the extra temporal entanglement in the symmetry broken phase may arise from the edge modes generated by the longitudinal field -- i.e., long-lived bound states of a domain-wall tied to the edge by a confining potential $\propto h$~\cite{Kormos17,LeroseQuasilocalized,RobinsonNonthermalStatesShort} -- generalizing the effect of edge modes in the integrable limit (cf. Sec.~\ref{sec4}).
Elucidating this intriguing issue is however beyond the scope of this Article, and is left to future investigations.

Using the numerically obtained MPS representation of the long-time IM, it is possible to fully access the non-integrable relaxation dynamics of a bulk subsystem over large time windows. To illustrate this, we computed the time-dependence of the dynamical correlation functions
\begin{align}
\label{eq_czz}
    C_{zz}(t) = \langle Z_0(t)Z_0(0)\rangle.
\end{align}
As reported in Fig.~\ref{fig_czz}, for this observable integrability breaking
leads to a slow decay to zero preceded by the fast initial decay characteristic
of the integrable limit. Deep in the paramagnetic phase, this observable shows
oscillating behavior, while this doesn't happen in the ferromagnetic phase.
The integrability breaking parameter affects the time at which the crossover
from fast to slow decay occurs, while the rate of the slow decay is nearly
independent of it.

\section{Conclusions and perspectives}\label{sec7} 

In this work, we used analytical methods and numerical computations to characterize the von Neumann entropy of the influence matrix, here dubbed  \textit{temporal entanglement} entropy. The possibility to efficiently simulate the quantum dynamics of arbitrary local subsystems of a many-body system crucially depends on the scaling of this quantity as a function of the system's evolution time.

We have established that TE entropy remains finite for arbitrarily long times (area-law scaling)
in a class of integrable Floquet quantum systems with underlying non-interacting quasiparticles. When integrability is weakly broken, TE entropy deviates from the area-law saturation at times which parametrically increase as the integrability-breaking perturbation is decreased. This allows us to efficiently simulate the long transient regime of integrability-breaking dynamics.

Another remarkable output of our analysis is the scaling behavior of TE as (Floquet) quantum critical points are approached.
In contrast to the ground-state spatial entanglement, which diverges logarithmically in one-dimensional critical systems, the infinite-temperature IM TE remains finite at critical points/lines, permitting efficient numerical simulations of dynamics uniformly in the phase diagram.
Signatures of criticality across phase transitions arise instead in the form of critical slowing down of TE saturation, and discontinuous jumps in its long-time saturation value. We have attributed this singular behavior to the changes in the spectral density, caused by the closing of the quasiparticle gap and by the appearance of long-lived edge coherence due to strong zero modes.
We note in passing that our work sheds light on the highly singular nature of the exactly solvable self-dual points of quantum circuits~\cite{Guhr1,Bertini2019} and on the structure of perturbation theory around them.

Our results naturally lend themselves to many interesting extensions and generalizations, which will lead to a complete picture of TE scaling using the ideas developed here.
First, our analysis directly applies to the continuous-time limit of Hamiltonian dynamics, by viewing the circuit as a Trotterization  $J\mapsto J \Delta t$, $g \mapsto g \Delta t$, $h \mapsto h \Delta t$, $t\mapsto t/\Delta t$ and taking the limit of vanishing Trotter step $\Delta t\to0$. The area law scaling found above implies that TE becomes \textit{vanishingly small} in the continuous limit (which corresponds to the bottom-left corner of Fig.~\ref{fig_integrable}a). This surprising behavior is further confirmed by our computations in several limiting regimes~\cite{SonnerAoP}. It can be ascribed to including the subsystem-environment interaction terms fully into the influence functional~\cite{FeynmanVernon}, in contrast to earlier tensor-network approaches where the splitting of interaction gates as two-site MPOs changes the scaling of the IM with the Trotter step~\cite{Banuls09,Chan21}.

Furthermore, our analytical results can be extended to more general unitary circuits such as brickwork structures with non-commuting interactions, and to non-stationary initial states.
These setups allow to use our approach to study global quenches and transport of conserved quantities.
In both cases, the generalization of the exact formula~\eqref{eq_IMoperator} for the influence matrix requires non-trivial technical steps compared to our derivation here, and will be reported elsewhere; we emphasize, however, that the results presented here are expected to be robust, since the qualitative structure of Eq.~\eqref{eq_IMoperator} only depends on generic quasiparticle properties, and nonequilibrium initial states are expected to modify the IM only near the temporal boundary $t=0$ (compared to the corresponding thermodynamic ensemble).

The approach developed here can also be readily applied to systems with quenched randomness and/or Hamiltonian noise. In particular, free-particle systems subject to on-site disorder exhibit persistent local interference effects due to Anderson localization. The IM approach allows to characterize this in terms of  TE scaling and patterns~\cite{ergodicitybreaking}, and to characterize the robustness of localization to introducing many-body interactions~\cite{Sonner20CharacterizingMBL}. 

An important question left open by our analysis concerns the ultimate fate of asymptotic TE entropy scaling for non-integrable systems in the long-time limit. Strongly chaotic quantum systems which induce rapid thermalization of their local subsystems, are expected to be characterized by influence actions that are quasi-local in time. Such IM, in addition to the quadratic part in Eq.~\eqref{eq_IMoperator}, include higher-order terms, in agreement with the weakly-interacting blip gas picture of Ref.~\cite{lerose2020}.
This quasi-locality in time may be expected to ultimately preserve the area-law scaling of TE entropy (cf. Fig.~\ref{fig_ent}), allowing to push efficient numerical simulations to arbitrarily long times and in the thermodynamic limit. From this perspective, the IM approach provides a systematic and non-perturbative means to go beyond exactly solvable maximally-chaotic models~\cite{NahumPRXOperatorSpreading,ChalkerPRX,BertiniSFF}, serving as a powerful approach to describe quantum thermalization. In the future work, we will aim to relate scaling of TE in non-integrable systems to thermalization properties of the system.

We finally point out that our work creates a bridge between the field of quantum many-body dynamics and the theory of open quantum systems, where influence functionals were first proposed~\cite{FeynmanVernon} and exploited~\cite{LeggettRMP,IFnanodevices}.
In this context, methods to describe open system dynamics based using tensor-networks have recently appeared, which effectively rely on the quasi-locality in time of the influence functional of non-interacting particle environments~\cite{MakriMakarov94,TEMPO,Cygorek21}.
Our work establishes that tensor-network description of non-interacting fermionic environments is efficient. We expect a further fruitful cross-fertilization of ideas between our approach to quantum dynamics and the research on open quantum systems (see also recent Ref.~\cite{Chan21}), in particular in describing complex baths of interacting particles and in adapting tools from open quantum system theory to better understand quantum thermalization.

\section{Acknowledgments}

This work was supported by the Swiss National Science Foundation and by the European Research Council (ERC) under the European Union's Horizon 2020 research and innovation programme (grant agreement No. 864597). 
We thank Bruno Bertini and Lorenzo Piroli for useful comments on the manuscript. Computations were performed at the University of Geneva on the ``Baobab'' and
``Yggdrasil'' HPC clusters.

\appendix

\section{\\Diagonalization of the open-boundary transverse-kicked Ising chain}

\label{app_diagonalizationKIC}

In this Appendix we report the derivation of the analytical solution of the semi-infinite
transverse-kicked Ising chain, i.e., the exact dynamics generated by the environment Floquet operator $F_E$ in the main text. This allows us to obtain explicit expressions for the quasiparticle spectrum $\{\phi_m\}$ and coefficients $\{\mathcal{C}_m\}$ on the boundary operator that couples to the subsystem, which characterize the interaction-picture unitary gates $\widetilde{F}_{\text{int}}(\tau)$ in Eq.~\eqref{eq_fermionicintpict}.

\subsection{Mapping to a linear Majorana map}
The model in Eq.~\eqref{eq_KIC} can be mapped to a quadratic model of fermions via the Jordan-Wigner transformation. Focusing on the right environment, composed of spins located on sites $j \ge 1$, we map
\begin{equation}
    \sigma^-_j
    = \prod_{i=0}^{j-1} e^{i\pi c^\dagger_i c_i} c^\dagger_j
    , \qquad
    X_j
    = 1-2c^\dagger_j c_j = e^{i\pi c^\dagger_j c_j},
\end{equation}
where $\sigma^{\pm}_j=\frac 1 2 (Y_j \pm i Z_j)$.
The operators $c_j$, $c_j^\dagger$ defined above satisfy the canonical fermionic algebra
\begin{equation}
\{ c_i, c_j \} = 0 , \qquad
\{ c_i, c^\dagger_j \} = \delta_{ij} .
\end{equation}
It is also convenient to introduce the real Majorana operators
 \begin{equation}
    a_{2j-1}=i(c_j^\dagger-c_j),
      \qquad
      a_{2j}=c_j+c_j^\dagger,
 \end{equation}
 with $\{a_m,a_n\}=2\delta_{mn}$.
In terms of this algebra, {the unitary gates in Eq.~\eqref{eq_KIC} become quadratic}, and the environment Floquet operator $F_E=U_2 U_1$ reads
\begin{equation}
\label{eq_fermionizedmodel}
{
\begin{split}
U_1 & =
\prod_{j=1}^\infty
e^{
i g (c_j c^\dagger_j-c^\dagger_jc_j)
}
=
\prod_{j=1}^\infty
e^{
 -g a_{2j-1} a_{2j}
}
\, ,
\\
U_2 & =
\prod_{j=1}^{\infty}
e^{
i J
(c^\dagger_j+c_j) (c_{j+1} -
 c^\dagger_{j+1})
}
=
\prod_{j=1}^{\infty}
e^{
- J
a_{2j} a_{2j+1}
}
\, .
\end{split}
}
\end{equation}
The interaction gate reads \begin{equation}
    F_{\text{int}}= e^{
i J
(c^\dagger_0+c_0) (c_{1} -
 c^\dagger_{1})
}
=
e^{- J a_0 a_1}
\, ,
\end{equation}
%
and we want to compute
the Heisenberg evolution of the environment boundary operator $i(c_{1} -
 c^\dagger_{1})=-a_1$, generated by periodic applications of the unitaries $U_1$ and $U_2$.
 Since the generators are quadratic, the Heisenberg evolution $a_i(\tau) \mapsto a_i(\tau+1)$ is linear: defining
 $a'_i=U_1^\dagger a_i U_1$,
 $a''_i=U_2^\dagger a'_i U_2$, we have, for $j\ge 1$,
 \begin{equation}
 \label{eq_Floquetmap}
 \begin{split}
 &
 \left\{
     \begin{split}
         a'_{2j-1} & = \cos(2g) \, a_{2j-1} - \sin(2 g) \, a_{2j} \,  , \\
         a'_{2j} & = \sin (2g) \, a_{2j-1} + \cos (2 g) \, a_{2j}\,   ,
     \end{split}
\right.
\\
&
\left\{
     \begin{split}
         a''_{2j} & = \cos(2J) \, a'_{2j} - \sin(2 J) \, a'_{2j+1 } \,  , \\
         a''_{2j+1} & = \sin (2J) \, a'_{2j} + \cos (2 J) \, a'_{2j+1} \, .
     \end{split}
\right.
\\
\end{split}
\end{equation}
The open boundary condition is imposed by setting $a''_1=a'_1$.

To solve for $a_1(\tau)$,
we diagonalize the composite Floquet map in Eq.~\eqref{eq_Floquetmap},
which is a (real) rotation.
We thus seek for a set of vectors $\{\psi^m_i\}_{i=1}^\infty$, labelled by the index $k$, that solve the linear system of equations $d_i''=e^{-i\phi_m} d_i$, $i=1,\dots,\infty$, for some $\phi_m\in[0,\pi)$.
 In this case, $\{(\psi^m_i)^*\}_{i=1}^\infty$ satisfies the same system with $\phi_m\mapsto-\phi_m$.
 From this solution we directly read off
  the quasienergy spectrum $\{\phi_m\}$, and the desired amplitudes $\{\mathcal{C}_m\equiv-\psi^m_1\}$ entering Eq.~\eqref{eq_fermionicintpict}.

\subsection{Bulk solutions}

Bulk translational invariance suggests the Bloch ansatz
\begin{equation}
    \begin{pmatrix}
    \psi_{2j-1} \\
    \psi_{2j}
    \end{pmatrix}
    = e^{ik(j-1)}
    \begin{pmatrix}
    \alpha \\
    \beta
    \end{pmatrix} \, .
\end{equation}
The eigenvector equation reduces to the $k$-dependent $2\times 2$ secular equation
\begin{equation}
\label{eq_Blochreduction}
    M_k \begin{pmatrix}
    \alpha \\
    \beta
    \end{pmatrix}
    =
    e^{i\phi_k}
    \begin{pmatrix}
    \alpha \\
    \beta
    \end{pmatrix}
\end{equation}
with
\begin{widetext}
\begin{equation}
    M_k= \begin{pmatrix}
    \cos(2J)\cos(2g)+ \sin(2J)\sin(2g) e^{-ik}
    &
    -\cos(2J)\sin(2g)+ \sin(2J)\cos(2g) e^{-ik}
    \\
    \cos(2J)\sin(2g)- \sin(2J)\cos(2g) e^{ik}
    &
    \cos(2J)\cos(2g)+ \sin(2J)\sin(2g) e^{ik}
    \end{pmatrix}
\end{equation}
\end{widetext}
Since $\det M_k\equiv 1$, from $\Tr M_k = 2\cos \phi_k$ we get
\begin{equation}
\label{eq_dispersionrel}
    \cos \phi_k =\cos(2J)\cos(2g)+ \sin(2J)\sin(2g) \cos k  \, .
\end{equation}
When $\phi_k$ or $-\phi_k$ lie in the interval $[\cos(2J\mp 2g),\cos(2J\pm 2g)]$, the wavevector $k$ is real.
In this case, the matrix $M_k$ can be parameterized as a $SU(2)$ rotation
\begin{equation}
    M_k = \cos \phi_k \mathbb{1} + i \sin \phi_k \; \hat n_k \cdot \vec{\sigma}
\end{equation}
[$\vec{\sigma}\equiv (\sigma^x,\sigma^y,\sigma^z)$ Pauli matrices], with a rotation angle $2\phi_k$ and a rotation axis
\begin{equation}
\label{eq_axis}
    \hat n_k =
    \frac{-1}{\sin\phi_k}
    \begin{pmatrix}
      \sin(2J)\cos(2g)\sin k
    \\
    \cos(2J)\sin(2g)-\sin(2J)\cos(2g)\cos k
    \\
    \sin(2J)\sin(2g)\sin k
    \end{pmatrix}
\end{equation}
($|\hat n_k|\equiv 1$).
Parametrizing $\hat n_k$ through standard polar and azimuthal angles
\begin{equation}
\label{eq_spherical}
   \hat n_k =
    \begin{pmatrix}
    \sin\eta_k \cos \xi_k
    \\
    \sin\eta_k \sin \xi_k
    \\
    \cos \eta_k
    \end{pmatrix} ,
\end{equation}
the eigenvectors $M_k \ket{\pm}_k = e^{\pm i \phi_k}\ket{\pm}_k$ can be presented in the form
\begin{equation}
\ket{+}_k=
    \begin{pmatrix}
      \cos \big( \frac{\eta_k}{2} \big)
    \\
    \sin \big( \frac{\eta_k}{2} \big)
    e^{i \xi_k}
    \end{pmatrix},
\quad
\ket{-}_k=
    \begin{pmatrix}
     - \sin \big( \frac{\eta_k}{2} \big)
    e^{-i \xi_k}
    \\
     \cos \big( \frac{\eta_k}{2} \big)
    \end{pmatrix}.
\end{equation}
Each quasienergy level $\phi_k\equiv\phi_{-k}$ is doubly degenerate by reflection symmetry, except $k=0$ and $k=\pi$.
Note that the symmetry relations $\xi_{-k}=\pi-\xi_k$, $\eta_{-k}=\pi-\eta_k$ imply $\ket{+}_{-k} \propto \ket{-}_k^*$, i.e., the reflection-symmetric solution coincides with the time-reversed one.

\subsection{Phase shift off the boundary}
For an infinite chain, these propagating solutions exhaust the spectrum.
The open boundary condition at $j=1$, however,
breaks the reflection degeneracy and constrains the relative phase $\rho_k \equiv e^{i\delta_k}$ in the superposition of incoming and outgoing waves,
\begin{equation}
\label{eq_eigenfunction}
    \begin{pmatrix}
    \psi^k_{2j-1} \\
    \psi^k_{2j}
    \end{pmatrix}
    = e^{-ik(j-1)}
    \ket{+}_{-k} +
    \rho_k \,
    e^{ik(j-1)}
    \ket{+}_{k}
\end{equation}
in such a way that
\begin{equation}
    e^{i\phi_k} \psi^k_1 \overset{!}{=} \cos(2g) \psi^k_1 - \sin(2g) \psi^k_2.
\end{equation}
This yields
\begin{equation}
     \rho_k=e^{i\delta_k}=
    \frac{[e^{i\phi_k}-\cos(2g)] \omega_k - \sin(2g)}{[e^{i\phi_k}-\cos(2g)]  + \sin(2g)\omega_k}
\end{equation}
where we defined $\omega_k \equiv \tan \frac{\eta_k}{2} \, e^{i\xi_k}$.
Since the eigenfunctions in Eq.~\eqref{eq_eigenfunction} are correctly normalized, we get
\begin{equation}
    \mathcal{C}_k = - \psi^k_1 = \sin \frac{\eta_k}{2} e^{-i\xi_k} - \cos \frac{\eta_k}{2} \, \rho_k \, .
\end{equation}

In the main text, the behavior of $\mathcal{C}_k$ as $k\to0$ and $\pi$ is crucial to obtain the long-time asymptotics of the function $\kappa(\tau)$ in Eq.~\eqref{eq_gamma}, and hence the area-law scaling of temporal entanglement entropy. It is straightforward to show that $\mathcal{C}_k$ vanishes in these limits: from Eqs.~\eqref{eq_spherical},~ \eqref{eq_axis} we find $\eta_0=\eta_\pi= \pi/2$, $\xi_0 =\xi_\pi = \pi/2$, which give $\rho_0=\rho_\pi=i$ and hence $\mathcal{C}_0=\mathcal{C}_\pi=0$. The non-vanishing coefficient of the linear term can be found by Taylor-expanding the expression of $\mathcal{C}_k$ in $k$ or $k-\pi$.

\subsection{Edge modes}
Open boundary conditions may induce additional non-propagating solutions to Eq.~\eqref{eq_Blochreduction} exponentially localized at the boundary, i.e.,
with purely imaginary wavevector, $\lambda\equiv e^{ik}=\pm e^{-\kappa}$, $\kappa>0$.

To find the edge modes and the conditions for their existence, we take on a dual approach, and view the eigenvector equations $\{ \psi_i'' = e^{i\phi} \psi_i\}$ for the linear Floquet map in Eqs.~\eqref{eq_Floquetmap} as a transfer-matrix construction of the solution starting from the boundary condition
\begin{equation}
    \begin{pmatrix}
    \psi^k_{1} \\
    \psi^k_{2}
    \end{pmatrix}
    =
    \begin{pmatrix}
    \alpha \\
    \beta
    \end{pmatrix}\equiv\ket{\psi_0}.
\end{equation}
In fact, the eigenvector equations may be cast in the recursive form
\begin{widetext}
\begin{equation}
    \begin{pmatrix}
    \psi_{2j+1} \\
    \psi_{2j+2}
    \end{pmatrix}
    =
    T_\phi
    \begin{pmatrix}
    \psi_{2j-1} \\
    \psi_{2j}
    \end{pmatrix}.
\end{equation}
A direct calculation shows
\begin{equation}
    T_\phi=
    \frac{1}{\sin(2J)}
    \begin{pmatrix}
    \sin(2g)e^{-i\phi}
    &
    \cos(2J)-\cos(2g)e^{-i\phi}
    \\
    \cos(2J)-\cos(2g)e^{-i\phi}
    &
    \quad\frac{e^{i\phi}}{\sin(2g)} - 2 \cos(2J) \cot(2g)
    + e^{-i\phi} \cos(2g)\cot(2g)
    \end{pmatrix}.
\end{equation}
This gives $\det T_\phi = 1$ and
\begin{equation}
    \frac{1}{2} \Tr T_\phi=
    \frac{1}{\sin (2J) \sin(2g)} [\cos\phi - \cos (2J)\cos (2g)],
\end{equation}
corresponding to the two dual eigenvalues $\lambda_\phi^{\pm 1} \equiv e^{\pm i k}$ connected with $\phi$ via the dispersion relation~\eqref{eq_dispersionrel}.
The eigenfunction thus takes the form
\begin{equation}
\label{eq_dualeigenfunction}
    \begin{pmatrix}
    \psi_{2j+1} \\
    \psi_{2j+2}
    \end{pmatrix}
    =
    T_\phi^j
    \ket{\psi_0}
    =
    \lambda_\phi^j
    \ket{+}_{\phi} \tensor[_\phi]{
    \braket{\widetilde{+}|\psi_0}
    }{}
    +
    \lambda_\phi^{-j}
    \ket{-}_{\phi} \tensor[_\phi]{
    \braket{\widetilde{-}|\psi_0}
    }{}
\end{equation}
where the tilde indicates that the bra and ket eigenvectors are not conjugate to each other when $\phi$ is outside the continuous quasienergy band~\eqref{eq_dispersionrel}.

The boundary condition takes the form
\begin{equation}
    \braket{b_\phi|\psi_0}=0
\end{equation}
with $\ket{b_\phi}\equiv(e^{i\phi}-\cos(2g),\sin(2g))^T$.
For $|\lambda_\phi|=1$, the boundary condition just determines the phase shift $e^{i\delta_k}$ as found in the previous subsection.
For $-1<\lambda_\phi<1$, which is our focus here, normalizability of the eigenfunction~\eqref{eq_dualeigenfunction} further requires $\tensor[_\phi]{
    \braket{\widetilde{-}|\psi_0}
    }{}\equiv 0$, i.e.,
$\ket{\psi_0} \propto \ket{+}_\phi$.
Edge modes exist if and only if there exists $\phi$ such that both these conditions are simultaneously satisfied.

Assuming $-1<\lambda_\phi<1$, direct substitution immediately gives the constraint $\phi=0,\pi$. Then, the value of $\lambda_{0,\pi}$ is determined by the dispersion relation, i.e.,
\begin{equation}
    \pm 1 = \cos(2J)\cos(2g)+ \sin(2J)\sin(2g)
    \frac {\lambda_{0,\pi}+\lambda_{0,\pi}^{-1}} {2}
    .
\end{equation}
Inversion of this equation gives
\begin{equation}
    \lambda_{0,\pi} =
    \frac {[\cos(2J)+1][-\cos(2g)\pm1]}
    {\sin(2J)\sin(2g)}
    =
    \left\{
    \begin{split}
        & \frac{\tan g}{\tan J}
        & \quad \text{for } \phi=0
        \\
        & - \frac{1}{\tan g\tan J}
        & \quad \text{for } \phi=\pi
        \\
    \end{split}
    \right.
    .
\end{equation}
\end{widetext}
Hence, in the quadrant $0\le J,g \le \pi/2$ analyzed in the main text, an edge mode with $\phi=0$ exists for $g<J$, and an edge mode with $\phi=\pi$ exists for $g>\pi/2-J$~\cite{DuttaPRB13}.
Their localization lengths $-1/\log|\lambda_{0,\pi}|$ drop to zero for $g=0,\pi/2$ or $J=\pi/2$. In these limiting cases, the edge modes can be understood by simple considerations, analogous to those in Ref.~\cite{Kitaev_2001}.

\section{\\Derivation of the exact influence matrix}

\label{app_derivationIM}

In this Appendix, we derive the exact formula in Eq.~\eqref{eq_IMoperator} for the IM of the transverse-field kicked Ising chain.

\subsection{Grassmann integral representation}

\label{app_grassmann}

The influence matrix in Eq. \eqref{eq_IMintpict} with the interaction-picture operators in Eq. \eqref{eq_fermionicintpict} is a partial trace of a sequence of exponentials of quadratic fermionic operators, over the Hilbert space spanned by the normal modes $\{ d_m,d^\dagger_m \}$ of the environment, as a function of the configuration of the fermion $c_0,c^\dagger_0$ at all times.
We compute this trace by means of its path integral representation, leading to a {Gaussian Grassmann integral}.

To obtain this representation, we 
insert a resolution of the identity by fermionic coherent states between each operator multiplication
in Eq.~\eqref{eq_IMintpict}  along the Schwinger-Keldysh closed-time contour. 
Integration is performed over the environment trajectories, treating the Grassmann variables associated with each input-output state of the subsystem trajectory as an external parameter.
This yields a (discrete-time) influence functional, expressed in the basis of fermionic coherent states.

Let us focus on the right environment, and assume  that it has finite length $L$.
Let us denote by $\eta,\bar\eta$ the Grassmann variables associated with the system and by $\bxi=(\xi_1,\dots,\xi_L)$, $\bar\bxi=(\bar\xi_1,\dots,\bar\xi_L)$  those associated with the quasiparticle modes of the environment.
The Grassmann path integral representation of Eq.~\eqref{eq_IMintpict} reads
\begin{widetext}
\begin{multline}
\label{eq_grassmannpathintegral}
\mathscr{I}[\eta^\pm_\tau,\bar\eta^\pm_\tau]
=
\frac{1}{2^L}
\int \Bigg[
\prod_{\tau=1}^t
d\bar\bxi^\pm_\tau d\bxi^\pm_\tau  \Bigg]
\; e^{-\bar\bxi^-_{t}\bar\bxi^+_t}
\\
\bigg(
\braket{
\bar\eta^+_t \bar\bxi^+_t |
\widetilde{F}_{\text{int}}(t) |
\eta^+_t \bxi^+_t
}
e^{\bxi^+_t\bar\bxi^+_{t-1}}
\braket{
\bar\eta^+_{t-1} \bar\bxi^+_{t-1} |
\widetilde{F}_{\text{int}}(t-1) |
\eta^+_{t-1} \bxi^+_{t-1}
}
\cdots \\ \cdots
\braket{ \bar\eta^+_1 \bar\bxi^+_1
| \widetilde{F}_{\text{int}}(1)
| \eta^+_1 \bxi^+_1
}
e^{\bxi^+_1\bar\bxi^-_1}
 \braket{
 \eta^-_1 \bxi^-_1 |
 \widetilde{F}_{\text{int}}^\dagger(1) |
 \bar\eta^-_1 \bar\bxi^-_1
 }
\cdots \\ \cdots
\braket{ \eta^-_{t-1} \bxi^-_{t-1}
| \widetilde{F}_{\text{int}}^\dagger(t-1)
| \bar\eta^-_{t-1} \bar\bxi^-_{t-1}
}
e^{\bar\bxi^-_{t-1}\bxi^-_t}
\braket{ \eta^-_t \bxi^-_t
| \widetilde{F}_{\text{int}}^\dagger(t)
| \bar\eta^-_t \bar\bxi^-_t
}
\bigg),
\end{multline}
where we took
$
    \rho_E = (1/2^L)\mathbb{1}_E
$ as in the main text, and the negative sign in the exponent of the first integrand term arises from the antiperiodic boundary condition prescription for the trace of a Grassmann kernel.
The representation \eqref{eq_grassmannpathintegral} is pictorially illustrated in Fig.~\ref{fig_grassmann}.

\begin{figure}[h]
    \centering
    \includegraphics[width=0.3\textwidth]{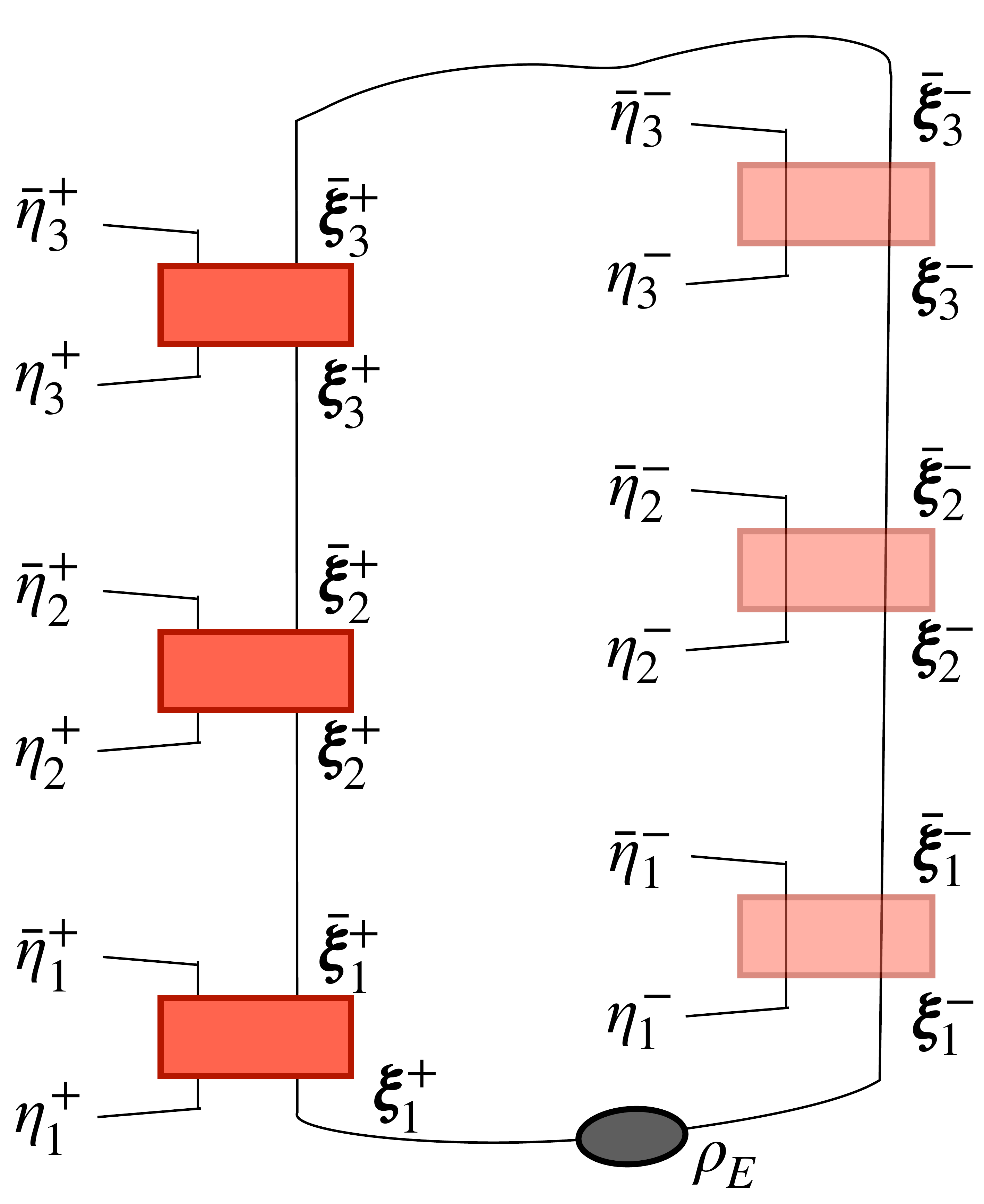}
    \caption{
    Illustration of the Grassmann path integral expression of the influence matrix in Eq.~\eqref{eq_grassmannpathintegral}. The wiggle in the top part of the contour represents the antiperiodic boundary condition prescription for the fermionic trace.
    }
    \label{fig_grassmann}
\end{figure}

Equation~\eqref{eq_grassmannpathintegral} is valid in full generality.
We specialize it to the integrable kicked Ising chain by
substituting Eq.~\eqref{eq_fermionicintpict}.
Using $e^{iJZ_0Z_1}=\cos J + i \sin J Z_0 Z_1$,
we find the conveniently factorized expression
\begin{equation}
\label{eq_Fintkernel}
\begin{split}
\braket{ \bar\eta,\bar\bxi
|
\widetilde{F}_{\text{int}}(\tau) |
\eta,\bxi
}
 & =
e^{\bar\eta \eta + \bar\bxi \bxi }
\bigg[ \cos J + \sin J \;
(\eta+\bar\eta)
\bigg(
\sum_{m=1}^L \mathcal{C}_k(\tau) \xi_m + \mathcal{C}^*_m(\tau) \bar\xi_m
\bigg)
\bigg] \\
& = \cos J \;
e^{\bar\eta \eta + \bar\bxi \bxi } \;
e^{ \tan J \; (\eta+\bar\eta)
\big(
\sum_{m=1}^L \mathcal{C}_m(\tau) \xi_m + \mathcal{C}^*_m(\tau) \bar\xi_m
\big) } \\
& =
\cos J \;
e^{\bar\eta \eta  } \;
\prod_{m=1}^L
e^{ \; \bar\xi_m \xi_m \; + \;  \tan J \; (\eta+\bar\eta)
\big(
\mathcal{C}_m(\tau) \xi_m + \mathcal{C}^*_m(\tau) \bar\xi_m
\big) }
\end{split}
\end{equation}
where we defined $\mathcal{C}_{m}(\tau) \equiv \mathcal{C}_{m}
e^{-i \tau \phi_m }$ for brevity.
Direct substitution into Eq.~\eqref{eq_grassmannpathintegral} gives
\begin{equation}
\label{eq_IMgrassmannintegral}
\mathscr{I}[\eta^\pm_\tau,\bar\eta^\pm_\tau]=
(\cos J)^{2t} \;
e^{\sum_{\tau=1}^t
\big(\bar\eta^+_{\tau} \eta^+_{\tau}
-
\bar\eta^-_{\tau} \eta^-_{\tau}
\big)
}
\prod_{m=1}^L
\frac 1 2  \mathcal{I}_m[\eta^\pm_\tau,\bar\eta^\pm_\tau] \, ,
\end{equation}
where the $m$-th single-mode influence matrix reads
\begin{multline}
\label{eq_singlemodegrassmannintegral}
\mathscr{I}_m[\eta^\pm_\tau,\bar\eta^\pm_\tau] =
\int \bigg[\prod_{\tau=1}^td\bar\xi^\pm_\tau d\xi^\pm_\tau\bigg]
\; e^{
\bar\xi^+_{t} \bar\xi^-_{t} +
\sum_{\tau=2}^{t}
\big(  (\bar\xi^+_{\tau}-\bar\xi^+_{\tau-1}) \xi^+_{\tau}
-
 (\bar\xi^-_{\tau}-\bar\xi^-_{\tau-1})
 \xi^-_{\tau}
\big)
+ \bar\xi^+_{1} \xi^+_{1}
- \bar\xi^-_{1} \xi^-_{1}
+ \xi^+_{1} \xi^-_{1}
}
 \\
 \times \;
e^{
\tan J \sum_{\tau=1}^t (\eta^+_\tau+\bar\eta^+_\tau)
\big(
\mathcal{C}_m(\tau) \xi^+_\tau + \mathcal{C}^*_m(\tau) \bar\xi^+_\tau
\big)
-
(\eta^-_\tau+\bar\eta^-_\tau)
\big(
\mathcal{C}^*_m(\tau) \xi^-_\tau + \mathcal{C}_m(\tau) \bar\xi^-_\tau
\big)
}
\end{multline}
(in the last expression we have dropped the label $m$ in the dummy integration variables $\xi_m$, $\bar\xi_m$).

We note that
the matrix elements $\mathscr{I}[\sigma^\pm_\tau,s^\pm_\tau]$ in the standard spin basis can be obtained by contracting the Grassmann IM in Eq.~\eqref{eq_grassmannpathintegral} with the Grassmann kernels of the operators $T_{\sigma,s}=\ket{\sigma} \bra{s}$.
To show this explicitly, we rewrite Eq.~\eqref{eq_IMintpict} as
\begin{multline}
\mathscr{I}[\sigma^\pm_\tau,s^\pm_\tau]
=
\Tr_{SE}
\Bigg[
\Big(\ket{s^-_t}\bra{s^+_t} \otimes \mathbb{1}_E \Big)
\widetilde{F}_{\text{int}}(t)
\Big(\ket{\sigma^+_{t}}\bra{s^+_{t-1}} \otimes \mathbb{1}_E \Big)
\cdots \\ \cdots
\Big( \ket{\sigma^+_{2}}\bra{s^+_1} \otimes \mathbb{1}_E \Big)
\widetilde{F}_{\text{int}}(1) \; \Big( \ket{\sigma^+_1}\bra{\sigma^-_1} \otimes \mathbb{1}_E \Big) \; \bigg(\mathbb{1}_S \otimes \rho_E \bigg) \;
\widetilde{F}_{\text{int}}^\dagger(1)
\Big( \ket{s^-_1}\bra{\sigma^-_2} \otimes \mathbb{1}_E \Big)
\cdots \\ \cdots
\Big( \ket{s^-_{t-1}}\bra{\sigma^-_t} \otimes \mathbb{1}_E \Big)
\widetilde{F}_{\text{int}}^\dagger(t)
\Bigg],
\end{multline}
\end{widetext}
where $S$ and $E$ denote the subsystem's and environment's Hilbert spaces, respectively.
To connect the spin matrix elements with the Grassmann path integral expression, we substitute the Grassmann kernels of the operators $T_{\sigma,s}\equiv \ket{\sigma}\bra{s} \otimes \mathbb{1}_E$ and $\widetilde{F}_{\text{int}}$. The latter is given by Eq.~\eqref{eq_Fintkernel} above, whereas the former reads
\begin{equation}
\label{eq_matrixelem}
\braket{ \eta,\bxi
| T_{\sigma,s}
| \bar\eta,\bar\bxi
}
\; = \;
 \frac 1 2 (1+i\sigma \eta)(1-i s \bar\eta) \; e^{\bxi \bar\bxi} \, .
\end{equation}
%
The polynomial in $\{\sigma^\pm,s^\pm\}$ arising from the matrix elements in Eq.~\eqref{eq_matrixelem} 
can be rigidly moved out of the integral preserving the time-ordering on the Keldysh contour.
The remaining path integral over $\bxi$, $\bar\bxi$ defines the Grassmann influence functional in Eq.~\eqref{eq_IMgrassmannintegral}.
Thus, its convolution with the polynomial in $\{\sigma^\pm,s^\pm\}$ allows to transform from the fermionic coherent-state basis to the original spin basis, as claimed.

\subsection{Integrating out the environment}
The path integral \eqref{eq_singlemodegrassmannintegral} can be evaluated exactly.
We use the {``complex'' Gaussian integral formula}
\newcommand{\bpsi}{\boldsymbol{\psi}}
\newcommand{\blam}{\boldsymbol{\lambda}}
\begin{equation}
\label{eq_gaussianformula}
I = \int \Big[\prod d\bar\psi d\psi\Big] e^{-\bar\bpsi A \bpsi} e^{ \bar\bpsi \blam - \bar\blam \bpsi } =
\det A \; e^{-\bar\blam A^{-1} \blam}
\end{equation}
where $\bpsi=(\psi_1,\dots,\psi_N)$, $\bar\bpsi=(\bar\psi_1,\dots,\bar\psi_N)$ are the integration Grassmann variables, and $\blam=(\lambda_1,\dots,\lambda_N)$, $\bar\blam=(\bar\lambda_1,\dots,\bar\lambda_N)$ are external Grassmann parameters.
This result is equivalent to {saddle-point integration}:
\begin{equation}
\begin{split}
S[\bar\bpsi,\bpsi] & = \bar\bpsi A \bpsi + \bar\blam \bpsi - \bar\bpsi \blam ,
\\
\frac{\partial S}{\partial \bar\bpsi}\bigg\rvert_{\bpsi^*} & = \frac{\partial S}{\partial \bpsi}\bigg\rvert_{\bar\bpsi^*} \overset{!}{=} 0
\\
\label{eq_saddlepoint}
I &= \int \Big[\prod d\bar\psi d\psi\Big] \; e^{-S[\bar\bpsi,\bpsi]}
 \\ &=
\det A \;
e^{-S[\bar\bpsi^*,\bpsi^*]}
\equiv
\det A \;
e^{-\bar\blam \bpsi^*} \, .
\end{split}
\end{equation}

To write Eq. \eqref{eq_singlemodegrassmannintegral} in the form \eqref{eq_gaussianformula}, we define the $2t$ pairs of Grassmann variables along the Keldysh contour:
\begin{equation}
\begin{split}
\bar\psi_\tau &= \left\{
\begin{split}
    \bar \xi^+_{\tau} & \qquad \tau=1,2,\dots,t \\
     \xi^-_{-\tau} & \qquad  \tau=-1,\dots,-t
\end{split}
\right.
, \\
\psi_\tau &= \left\{
\begin{split}
     \xi^+_{\tau} & \qquad \tau=1,2,\dots,t \\
     \bar \xi^-_{-\tau} & \qquad  \tau=-1,\dots,-t
\end{split}
\right.
\end{split}
\end{equation}
With these definitions, the action in Eq.~\eqref{eq_singlemodegrassmannintegral} only pairs $\bar\bpsi$ and $\bpsi$.
The result of the integration is thus Eq.~\eqref{eq_saddlepoint}, provided we identify $-\bar\blam$ with the array of coefficients of $\bpsi$, and provided we solve the saddle point equation for $\bpsi^*$.
The coefficients of $\xi^+_\tau$ and $\bar\xi^-_\tau$ are
\begin{equation}
\label{eq_linearshift}
\begin{split}
-\bar\lambda_{\tau} & =
\tan J \; (\eta^+_\tau + \bar\eta^+_\tau) \;  \mathcal{C}_k(\tau) ,
\\
-\bar\lambda_{-\tau} & = - \tan J \; (\eta^-_\tau + \bar\eta^-_\tau) \;  \mathcal{C}_k(\tau) ,
\end{split}
\end{equation}
respectively.
The saddle point equations read:
\begin{equation}
\begin{split}
\frac{\partial S}{\partial \bar\xi^+_\tau} &= \xi^+_{\tau+1}-\xi^+_{\tau} + \tan J \; (\eta^+_\tau + \bar\eta^+_\tau) \;  \mathcal{C}^*_k(\tau) =0
\\
\frac{\partial S}{\partial \xi^-_\tau} &= \bar\xi^-_{\tau-1}-\bar\xi^-_{\tau}
- \tan J \; (\eta^-_\tau + \bar\eta^-_\tau) \;  \mathcal{C}^*_k(\tau) =0
\\
\frac{\partial S}{\partial \xi^-_1} &=
\xi^+_{1} -\bar\xi^-_1 -
 \tan J \; (\eta^-_1 + \bar\eta^-_1) \;  \mathcal{C}^*_k(1)  =0
\\
\frac{\partial S}{\partial \bar\xi^+_t} &=
- \bar\xi^-_t -
\xi^+_{t}
+ \tan J \; (\eta^+_t + \bar\eta^+_t) \;  \mathcal{C}^*_k(t)  =0
\end{split}
\end{equation}
From the first and second equation we find the preliminary expressions
\begin{equation}
\label{eq_preliminaryspsol}
\begin{split}
\xi^+_\tau & = \xi^+_1 - \tan J \sum_{s=1}^{\tau-1} (\eta^+_s + \bar\eta^+_s) \;  \mathcal{C}^*_k(s)
\\
\bar\xi^-_\tau &= \bar\xi^-_t + \tan J \sum_{s=\tau+1}^{t} (\eta^-_s + \bar\eta^-_s) \;  \mathcal{C}^*_k(s)
\end{split}.
\end{equation}
Setting $\tau=t$ in the first and $\tau=1$ in the second, summing the two and substituting the third equation above on the l.-h.s., we get
\begin{multline}
\bar\xi^-_t -
\xi^+_{t}
+
\tan J \; (\eta^+_1 + \bar\eta^+_1) \;  \mathcal{C}^*_k(1)
 \\+
\tan J
\sum_{s=1}^{t-1}
( \eta^-_s + \bar\eta^-_s - \eta^+_s - \bar\eta^+_s) \mathcal{C}^*_k(s)
=0.
\end{multline}
Exploiting now the fourth saddle-point equation above, we determine $\bar\xi^-_t$,
$\xi^+_{t} $ and hence $\bar\xi^-_1$,
$\xi^+_{1} $:
\begin{equation}
\begin{split}
\xi^+_1 & = \frac{\tan J}{2} \sum_{s=1}^t
( \eta^-_s + \bar\eta^-_s + \eta^+_s + \bar\eta^+_s) \mathcal{C}^*_k(s) \\
\bar\xi^-_t & = \frac{\tan J}{2} \sum_{s=1}^t
( \eta^+_s + \bar\eta^+_s - \eta^-_s - \bar\eta^-_s) \mathcal{C}^*_k(s)
\end{split}
\end{equation}
Hence, from Eqs. \eqref{eq_preliminaryspsol} we finally arrive at the saddle-point solution
\begin{widetext}
\begin{equation}
\label{eq_spsol}
\begin{split}
\big(\xi^+_\tau\big)^* & = \frac{\tan J}{2}
\bigg[\sum_{s=1}^{\tau-1}
( \eta^-_s + \bar\eta^-_s - \eta^+_s - \bar\eta^+_s) \mathcal{C}^*_k(s)
+ \sum_{s=\tau}^{t}
( \eta^-_s + \bar\eta^-_s + \eta^+_s + \bar\eta^+_s) \mathcal{C}^*_k(s)
\bigg],
\\
\big(\bar\xi^-_\tau\big)^* & = \frac{\tan J}{2}
\bigg[\sum_{s=1}^{\tau}
( \eta^+_s + \bar\eta^+_s - \eta^-_s - \bar\eta^-_s) \mathcal{C}^*_k(s)
+ \sum_{s=\tau+1}^{t}
( \eta^-_s + \bar\eta^-_s + \eta^+_s + \bar\eta^+_s) \mathcal{C}^*_k(s)
\bigg].
\end{split}
\end{equation}

The result of the integration of the quasiparticle mode $k$ is thus computed by substituting the coefficients \eqref{eq_linearshift} and the saddle-point solution \eqref{eq_spsol} into the Gaussian integral formula \eqref{eq_saddlepoint}.
Introducing the convenient variables
\begin{equation}
\label{eq_zetatheta}
\left\{
\begin{split}
\zeta^\pm_\tau &= \frac{1} {\sqrt{2}} (\eta^\pm_\tau + \bar\eta^\pm_\tau),
\\
\bar\zeta^\pm_\tau &= \frac{1} {\sqrt{2}} (\eta^\pm_\tau - \bar\eta^\pm_\tau)
\end{split}
\right.
\end{equation}
and rearranging the terms, we find
\begin{equation}
\frac{\mathscr{I}_m[\zeta^\pm_\tau,\bar\zeta^\pm_\tau]}{\det A} =
\exp\bigg\{
 2(\tan J)^2
 \bigg[
\sum_{\tau,s=1}^t
\Re[\mathcal{C}_m(\tau)\mathcal{C}^*_m(s)] \; \zeta^+_\tau \zeta^-_s
+
\sum_{\tau<s}^t
\Re[\mathcal{C}_m(\tau)\mathcal{C}^*_m(s)] \;
\big( \zeta^+_\tau \zeta^+_s
-  \zeta^-_\tau \zeta^-_s \big)
\bigg]
\bigg\}
\end{equation}
Plugging this into Eq. \eqref{eq_IMgrassmannintegral} we find
\begin{equation}
\mathscr{I}[\zeta^\pm_\tau,\bar\zeta^\pm_\tau]=
(\cos J)^{2t} \;
e^{\sum_{\tau=1}^t
\big(\zeta^+_{\tau} \bar\zeta^+_{\tau}
-
\zeta^-_{\tau} \bar\zeta^-_{\tau}
\big)
}
\bigg(\frac{\det A}{2}\bigg)^L
e^{
 \Big[
\sum_{\tau,s}
\kappa(\tau-s) \; \zeta^+_\tau \zeta^-_s
+
\sum_{\tau<s}
\kappa(\tau-s) \;
\big( \zeta^+_\tau \zeta^+_s
-  \zeta^-_\tau \zeta^-_s \big)
\Big]
}
\end{equation}
where we have defined the real function
\begin{equation}
\kappa(\tau-s) =
2(\tan J)^2 \sum_{m=1}^L
\Re[\mathcal{C}_m(\tau)\mathcal{C}^*_m(s)]
=
2(\tan J)^2 \sum_{m=1}^L
\big\lvert \mathcal{C}_{m} \big\rvert^2 \cos[\phi_m (\tau-s)] \, .
\end{equation}
We note that  the matrix $A$ in the action is independent of the subsystem-environment coupling $J$.
Since for $J=0$ the trace must equal $1$, we find $\det A=2$ (which can be easily checked explicitly). Thus, we obtain the final result for the {general influence matrix of an integrable kicked Ising chain}:
\begin{equation}
\label{eq_grassmannIM}
\mathscr{I}[\zeta^\pm_\tau,\bar\zeta^\pm_\tau]=
(\cos J)^{2t} \;
e^{\sum_{\tau}
\big(\zeta^+_{\tau} \bar\zeta^+_{\tau}
-
\zeta^-_{\tau} \bar\zeta^-_{\tau}
\big)
}
\exp
 \bigg[
\sum_{\tau,s}
\kappa(\tau-s) \; \zeta^+_\tau \zeta^-_s
+
\sum_{\tau<s}
\kappa(\tau-s) \;
\big( \zeta^+_\tau \zeta^+_s
-  \zeta^-_\tau \zeta^-_s \big)
\bigg]
.
\end{equation}
Equation~\eqref{eq_IMoperator} of the main text directly follows upon translating this wavefunction into the familiar operator language (the non-barred [barred] variables become the ``$\uparrow$'' [``$\downarrow$''] creation operators).


In Eq.~\eqref{eq_BCSform} of the main text, we use the expression of the IM wavefunction with Keldysh-rotated fields:
Defining
\begin{equation}
\label{eq_Keldyshrotation}
\left\{
\begin{split}
\zeta^{cl,q}_\tau &= \frac 1 {\sqrt{2}} (\zeta^+_\tau \pm \zeta^-_\tau),
\\
\bar\zeta^{cl,q}_\tau &= \frac 1 {\sqrt{2}} (\bar\zeta^+_\tau \pm \bar\zeta^-_\tau),
\end{split}
\right.
\end{equation}
we get
\begin{equation}
\label{eq_grassmannIMkeldysh}
{
\mathscr{I}\big[\zeta^{cl,q}_\tau \, , \, \bar\zeta^{cl,q}_\tau\big]=
(\cos J)^{2t} \;
\exp\bigg[
\sum_{\tau}
\big(\zeta^{cl}_{\tau} \bar\zeta^q_{\tau}
+
\zeta^q_{\tau} \bar\zeta^{cl}_{\tau}
\big)
+ 
2 \sum_{\tau,s}
\Theta(s-\tau)\kappa(s-\tau) \; \zeta^q_\tau \zeta^{cl}_s
\bigg]
}
\end{equation}
where $\Theta(x)=[1+\sign(x)]/2$ is Heaviside's theta function [$\Theta(0)\equiv 1/2$].
\end{widetext}

\bibliography{mbl}

\end{document}